\documentclass[pinchcr,eqsecnum]{revtex4}
\usepackage{amsfonts}
\usepackage{graphicx}
\usepackage{epsfig}
\def\bea{\begin{eqnarray}}
\def\eea{\end{eqnarray}}
\def\be{\begin{equation}}
\def\ee{\end{equation}}

\begin{document}
\title{Quantum Impurity Problems in Condensed Matter Physics\cite{houches}}
\affiliation{Department of Physics and Astronomy, University of British 
Columbia, Vancouver, B.C., Canada, V6T 1Z1}
\begin{abstract}
Impurities are ubiquitous in condensed matter.  Boundary Conformal Field Theory (BCFT)
provides a powerful method to study a localized quantum impurity interacting 
with a gapless continuum of excitations. The results can also be implied 
to nanoscopic devices like quantum dots. In these lecture notes, I review 
this field, including the following topics:

I. General Renormalization Group (RG) framework for quantum impurity problems: example
of simplest Kondo model

II. Multi-channel Kondo model

III. Quantum Dots: experimental realizations of one and two channel Kondo models

IV. Impurities in Luttinger liquids: point contact in a quantum wire

V. Quantum impurity entanglement entropy

VI. Y-junctions of Luttinger liquids

VII. Boundary condition changing operators and the X-ray edge problem
\end{abstract}
\maketitle
\section{Quantum Impurity Problems and the Renormalization Group}
A remarkable property of nature, that has intriqued physicists for many years, 
is universality at critical points.\cite{Ma} An impressive example is the critical 
point of water. By adjusting the temperature and pressure, a critical 
point is reached where the correlation length diverges and the long 
distance physics becomes the same as that of the Ising model. A microscopic 
description of water is very complicated and bears very little connection 
with the Ising model; in particular, there is no lattice, no spin operators 
and not even any $Z_2$ symmetry. Nonetheless, various experimentally 
measured critical exponents appear to be exactly the same as those of 
the Ising model. Furthermore, the best description of this universal long 
distance behaviour is probably provided by the  $\varphi^4$ field theory at 
its critical point. Our understanding of universality is based upon 
the RG. For a system at or near a critical point with a diverging 
correlation length, it is convenient to consider an 
effective free energy (or Hamiltonian), used only to describe long distance properties, 
which is obtained by integrating out short distance 
degrees of freedom.  It is found that the same long-distance 
Hamiltonian, characterizing an RG fixed point, is obtained 
from many different microscopic models. These fixed point 
Hamiltonians are universal attractors for all microscopic models.\cite{Ma}

The same features hold for quantum models of many body systems 
at low temperature. In many cases such models can exhibit 
infinite correlation lengths and vanishing excitation energy 
gaps.  In this situation one again expects universality. Important 
examples are the Fermi liquid fixed point for interacting 
electrons in 3 dimensions (D=3),\cite{Shankar} its cousin, the Luttinger 
liquid fixed point in D=1 and various models of interacting quantum spins.\cite{Affleck0,Giamarchi}

In these lectures, I will be concerned with a single quantum impurity 
embedded in such a critical system. The quantum impurity can 
be of quite a general form, possibly comprising several 
nearby impurities.  If we study its behaviour at long length 
scales (compared to all microscopic lengths including the spatial 
extent of the impurity and the range of its interactions with the host) 
and at low energies compared to all microscopic energy scales, then 
universality again emerges. These single impurity models, while 
simplified, have the attractive feature that such powerful 
methods as BCFT can be used to tackle them. They provide 
quite non-trivial examples of quantum critical phenomena 
and, in some cases, appear to be good descriptions of experimental reality. 

An important example is provided by the simplest version of the Kondo model.\cite{Kondo,Hewson,Affleckrev}
This is a model invented to describe a single magnetic impurity 
(such as an iron atom) in a non-magnetic metal (such as copper). 
Traditional experiments in this field always involve a finite 
density of impurities, but if this density is low enough, 
we may consider only one of them; technically this gives 
the first term in a virial expansion in the impurity density. 
Furthermore, these models can be applied to situations 
where the single impurity is a nanostructure device like a quantum dot. 
A simple Hamiltonian to describe this system can be written:
\be H=\int d^3k \psi^{\dagger\alpha}_{\vec{k}}
\psi_{\vec{k}\alpha}\epsilon(k)+J \int {d^3kd^3k'\over (2\pi )^3}
\psi^{\dagger \alpha}_{\vec k} \frac{\vec\sigma^\beta_\alpha }{2}\psi_{\vec{k'}\beta}
\cdot \vec S
\ee
Here $\psi_{\vec k\alpha}$ annihilates an electron of wave-vector $\vec k$ 
and spin $\alpha$ and is normalized so that:
\be \{\psi^{\dagger \alpha}_{\vec k},\psi_{\vec k'\beta}\}=\delta^{\alpha}_{\beta}\delta^3(\vec k-\vec k').\ee
Repeated spin indices are summed over. 
  $\epsilon(\vec k)$ is the dispersion relation 
for the electrons, which we will usually approximate by the free electron 
form: 
\be \epsilon (\vec k)={k^2\over 2m}-\epsilon_F\ee
where $\epsilon_F$ is the Fermi energy.  (So this 
is actually $H-\mu \hat N$.) $\vec S$ is an impurity 
spin operator, of magnitude $S$. $J$ measures the strength 
of a Heisenberg type exchange interaction between the 
electron spin density and the impurity spin. Usually $J>0$.  
Note that this form 
of interaction is a $\delta$-function in position space:
\be H=\int d^3r\left[\psi^{\dagger}(\vec r)\left(-{\nabla^2\over 2m}-\epsilon_F \right) \psi (\vec r)
+J\delta^3(r)\psi^\dagger {\vec \sigma \over 2}\psi \cdot \vec S\right].\label{Hps}\ee
Here we have suppressed the spin indices completely. Actually, this model is ultra-violet divergent unless we truncate the integral over $\vec k$, $\vec k'$ 
in the interaction term.  Such a truncation is assumed here but its details will not be important 
in what follows.
The dimensionless measure of the strength of the Kondo interaction is 
\be \lambda \equiv J\nu,\ee
where $\nu$ is the density of states, per unit energy per unit volume per spin.  For free electrons this is:
\be \nu = {mk_F\over \pi^2}\ee
where $k_F$ is the Fermi wave-vector. 
Typically, $\lambda \ll 1$. 

This model is a considerable simplification of reality. In particular, 
electrons in metals interact with each other via the Coulomb interaction 
and this is neglected. This can be justified using Fermi liquid ideas. 
Since $\lambda \ll 1$, the Kondo interaction only affects electrons 
close to the Fermi surface. The Coulomb interactions become increasingly 
ineffective for these electrons, as can be seen from phase space arguments, 
after taking into account screening of the Coulomb interactions. The free 
electron Hamiltonian (with an appropriate effective mass) represents 
the fixed point Hamiltonian, valid at low energies. Treating the 
Kondo interaction as a $\delta$-function is another approximation; 
a more realistic model would give it a finite range. Again, if we 
are concerned with the long distance, low energy physics, we might 
expect this distinction to be unimportant. The spherical symmetry 
of the dispersion relation and Kondo interaction will considerably simplify
our analysis, but again can be seen to be inessential. 

Due to the absence of bulk interactions, the $\delta$-function form of the 
Kondo interaction and the spherical symmetry of $\epsilon (k)$, we 
may usefully expand the electron operators in spherical harmonics, finding 
that only the s-wave harmonic interacts with the impurity.  (See, for example, 
Appendix A of [\onlinecite{Affleck3}].) This gives 
us an effectively one-dimensional problem, defined on the half-line, $r>0$, 
with the impurity sitting at the beginning of the line, $r=0$. Thus we write:
\be \psi_{\vec k}={1\over \sqrt{4\pi}k}\psi_0(k)+\ \ \hbox{higher harmonics}.\ee
Next we restrict the integral over $k$ in the Hamiltonian to a narrow band around 
the Fermi wave-vector:
\be -\Lambda <k-k_F<\Lambda .\ee
This is justified by the fact that $\lambda \ll 1$. To be more accurate, 
we should integrate out the Fourier modes further from the Fermi surface, 
renormalizing the Hamiltonian in the process.  However, for small $\lambda$ 
this only generates small corrections to $H$ which we simply ignore.  
Actually, this statement is only true if $\Lambda$ is chosen to be small 
but not {\it too} small. We want it to be $\ll k_F$.  However, 
if it becomes arbitrarily small, eventually the renormalized $\lambda$ 
starts to blow up, as we discuss below. Thus, we assume that $\Lambda$ is chosen judiciously 
to have an intermediate value. We can then approximate the dispersion relation by:
\be \epsilon (k)\approx v_F(k-k_F).\ee
We now define the following position space fields:
\be \psi_{L/R}(r)\equiv \int_{-\Lambda}^\Lambda dke^{\pm ikr}\psi_0(k_F+k).\ee
Note that these obey the boundary condition:
\be \psi_L(t,r=0)=\psi_R(t,r=0).\label{freebc}\ee
Furthermore, they obey approximately the anti-commutation relations:
\be \{\psi_{L/R}(x),\psi_{L/R}^\dagger (x')\}=2\pi \delta (x-x').\label{1Dnorm}\ee
[This is only approximately true at long distances.  The Dirac $\delta$-function 
is actually smeared over a distance of order $1/\Lambda$.  Note also 
the unconventional normalization in Eq. (\ref{1Dnorm}).]
The Hamiltonian can then be written:
\be H= {v_F\over 2\pi }i\int_0^\infty dr\left(\psi_L^\dagger {d\over dr}\psi_L-\psi_R^\dagger {d\over dr}\psi_R
\right)+v_F\lambda \psi_L^\dagger (0){\vec \sigma \over 2}\psi_L(0)\cdot \vec S +\hbox{higher harmonics}.
\label{Hb}\ee
The ``higher harmonics'' terms in $H$ are non-interacting and we will generally ignore them. 
This is a (1+1) dimensional massless Dirac fermion (with 2 ``flavours'' or spin components) defined 
on a half-line interacting with the impurity spin. The velocity of light 
is replaced by the Fermi velocity, $v_F$. We shall generally set $v_F=1$. Note that in a space-imaginary time 
representation, the model is defined on the half plane and there is an interaction 
with the impurity spin along the edge, $r=0$. Since the massless Dirac fermion model 
is a conformal field theory, this is a type of conformal field theory (CFT) with a boundary. 
However, it is a much more complicated boundary than discussed in John Cardy's lectures 
at this summer school.  There he considered CFT's with conformally invariant boundary 
conditions. If we set $\lambda =0$ then we have such a model since 
the boundary condition (BC) of Eq. (\ref{freebc}) is conformally invariant. More precisely, 
we have a boundary conformal field theory (BCFT) and a decoupled spin, sitting at the boundary. 
 However, for $\lambda \neq 0$, we do not have merely a BC but a 
boundary interaction with an impurity degree of freedom. Nonetheless, as we shall see, 
the low energy fixed point Hamiltonian is just a standard BCFT. 

Although the form of the Hamiltonian in Eq. (\ref{Hb}) makes the connection 
with BCFT most explicit, it is often convenient to make an ``unfolding'' transformation. 
Since $\psi_L(t,x)$ is a function of $(t+x)$ only and $\psi_R(t,x)$ is a function of 
$(t-x)$ only, the  boundary condition, of Eq. (\ref{freebc}) implies that 
we may think of $\psi_R$ as the analytic continuation of $\psi_L$ to 
the negative $r$ axis:
\be \psi_R(r)=\psi_L(-r),\ \  (r>0)\label{fold}\ee
and the Hamiltonian can be rewritten:
\be H= {v_F\over 2\pi }i\int_{-\infty}^\infty dr \psi_L^\dagger {d\over dr}\psi_L
+v_F\lambda \psi_L^\dagger (0){\vec \sigma \over 2}\psi_L(0)\cdot \vec S .
\label{HL}\ee
We have reflected the outgoing wave to the negative $r$-axis. In this 
representation, the electrons move to the left only, interacting with 
the impurity spin as they pass the origin. 

The phrase ``Kondo problem'', as far as I know, refers to the infrared divergent 
property of perturbation theory, in $\lambda$, discovered by Kondo in the mid-1960's. 
In the more modern language of the RG, this simply means that the renormalized 
coupling constant, $\lambda (E)$, increases as the characteristic energy scale, $E$, 
is lowered. The ``problem'' is how to understand the low energy behaviour 
given this failure of  perturbation theory, a failure which occurs despite 
the fact that the original coupling constant $\lambda \ll 1$. The $\beta$-function 
may be calculated using Feynman diagram methods; the first few diagrams are 
shown in Fig. (\ref{fig:pert}).  The dotted line represents the impurity spin. 
The simplest way to deal with it is to use time-ordered real-time perturbation 
theory and to explicitly evaluate the quantities:
\be {\cal T}<0|S^a(t_1)S^b(t_2)S^c(t_3)\ldots |0>.\ee
(For a detailed discussion of this approach and some third order calculations, 
see, for example, [\onlinecite{Barzykin}].) 
Since the non-interacting part of the Hamiltonian is independent of $\vec S$, 
these products are actually time-independent, up to some minus signs 
arising from the time-ordering. For instance, for the $S=1/2$ case:
\be {\cal T}<0|S^x(t_1)S^y(t_2)|0>=\theta (t_1-t_2)S^xS^y+\theta (t_2-t_1)S^yS^x
=\hbox{sgn} (t_1-t_2)iS^z.\ee
Here $\hbox{sgn} (t)$ is the sign function, $\pm 1$ for $t$ positive or negative respectively. 
Using 
the spin commutation relations and $\vec S^2=S(S+1)$ it is possible to explicitly 
evaluate the expectation values of any of the spin products occuring in perturbation theory. 
We are then left with standard fermion propogators. These are simplified by 
the fact that all fermion fields occuring in the Kondo interaction are at $r=0$; we 
suppress the spatial labels in what follows. 
For instance, the second order diagram is:
\be -\frac{\lambda^2}{2}\int  dt\ dt'T(S^a(t)S^b(t'))\cdot
T[\psi^\dagger (t)\frac{\sigma^a}{2}\psi (t)\psi^\dagger
(t')\frac{\sigma^b}{2} \psi (t')],\ee
which can be reduced, using Wick's theorem to:
\be \frac{\lambda^2}{2}\int  dt\
dt'\psi^\dagger \frac{\vec{\sigma}}{2}\psi\cdot\vec{S}\
\hbox{sgn}(t-t') <0|\psi(t)\psi^\dagger(t')|0>\label{tint}
.\ee
The free fermion propogator is simply:
\be G(t)={-i\over t}.\ee
This gives a correction to the effective coupling constant:
\be \delta \lambda = {\lambda^2\over 2}\int dt {\hbox{sgn}(t)\over t}\label{tint2}\ee
Integrating symmetrically, Eq. (\ref{tint}) would give 
zero if the factor sgn $(t)$, coming from the impurity spin Green's function, were absent. 
This corresponds to a cancellation between particle and hole contributions. 
This is as it should be. If we have a simple non-magnetic scatterer, with no 
dynamical degrees of freedom, there is no renormalization. The Kondo 
problem arises entirely from the essentially quantum-mechanical nature of the impurity spin. 
Including the sgn $(t)$ factor, the integral in Eq. (\ref{tint}) is infrared 
and ultra-violet log-divergent. In an RG transformation, we only integrate over a restricted 
range of wave-vectors, integrating out modes with $D'<|k|<D$. We then obtain:
\be \delta \lambda = \lambda^2\ln (D/D').\ee
The corresponding $\beta$-function, to this order is then:
\be {d\lambda\over d\ln D}=-\lambda^2 +\ldots \ee
Solving for the effective coupling at scale $D$, in terms of its bare value $\lambda_0$ 
at scale $D_0$ we obtain:
\be \lambda (D)\approx {\lambda_0\over 1-\lambda_0\ln (D_0/D)}.\ee
If the bare coupling is {\it ferromagnetic}, $\lambda_0<0$, then $\lambda (D)$ 
is well-behaved, getting smaller in magnitude
 at lower energy scales.  However, if it is antiferromagnetic, 
$\lambda (D)$ continues to increase as we reduce the energy scale until it gets 
so large that lowest order perturbation theory for the $\beta$-function breaks down. 
We may estimate the energy scale where this happens as:
\be T_K\approx D_0\exp (-1/\lambda_0).\ee
The scale $D_0$, which plays the role of the ultra-violet cut-off, is of order 
the band width or Fermi energy and $D_0=v_F/\Lambda$ where $\Lambda$ is the cut-off 
in momentum units. 

After many years of research by many theorists, a very simple picture immerged 
for the low energy behaviour of the Kondo model, due in large part to 
the contributions of PW Anderson,\cite{Anderson}
 K Wilson,\cite{Wilson} P Nozi\`eres\cite{Nozieres1} and collaborators. We may 
think of $\lambda$ as renormalizing to $\infty$.  What is perhaps 
surprising is that the infinite coupling limit is actually very simple. 
To see this, it is very convenient to consider a lattice model, 
\be H=-t\sum_{i=0}^\infty (\psi^\dagger_i\psi_{i+1}+\psi^\dagger_{i+1}
\psi_i)+J\psi^\dagger_0
\frac{\vec\sigma}{2}\psi_0\cdot \vec S .\label{lattice}\ee
 The strong coupling limit corresponds to $J\gg t$. 
It is quite easy to solve this limit exactly. One electron sits at site 0 and 
forms a spin singlet with the impurity, which I assume to have $S=1/2$ for now. 
$|\Uparrow\downarrow >-|\Downarrow\uparrow >$.  (Here the double arrow 
refers to the impurity spin and the single arrow to the spin of the electron 
at site zero.) The other electrons can do anything they like, as long 
as they don't go to site 0. Thus, we say the impurity spin is ``screened'', 
or more accurately has formed a spin singlet. To understand the low energy 
effective Hamiltonian, we are more interested in what the other electrons are doing, 
on the other sites. If we now consider a small but finite $t/J$, the other 
electrons will form the usual free fermion ground state, filling a Fermi sea, 
but with a modified boundary condition that they cannot enter site 0. 
It is as if there were an infinite repulsion at site 0. The single 
particle wave-functions are changed from $\sin k(j+1)$ to $\sin kj$. 
In the particle-hole (PH) symmetric case of a half-filled band, $k_F=\pi /2$, 
the phase shift at the Fermi surface is $\pi /2$. In this one-dimensional 
case, we take the continuum limit by writing:
\be \psi_j\approx e^{ik_Fj}\psi_R(j)+e^{-ik_Fj}\psi_L(j).\ee
For $\lambda =0$, in the PH symmetric case, the open boundary condition for the lattice model 
corresponds to
\be \psi_L(0)=\psi_R(0)\ee
in the continuum model, just as in D=3. On the other hand, the strong coupling BC is:
\be \psi_L(0)=-\psi_R(0).\label{BCs}\ee
The strong coupling fixed point is
the same as the weak coupling fixed point except for a change in
boundary conditions (and the removal of the  impurity).  We 
describe the strong coupling fixed point by the conformally invariant 
BC of Eq.  (\ref{BCs}). 

This simple example illustrates the main ideas of the BCFT approach to 
quantum impurity problems. In general, we consider systems 
whose long-distance, low energy behaviour, in the absence of 
any impurities, is described by a CFT. Examples include 
non-interacting fermions in any dimension, or interacting fermions 
(Luttinger liquids) in D=1. We then add some interactions, 
involving impurity degrees of freedom, localized near $r=0$. 
Despite the complicated, interacting nature 
of the boundary in the microscopic model, the low energy long 
distance physics is always described by a conformally invariant BC. 
The impurity degrees of freedom always either get screened or decouple, 
or some combination of both. Why should this be true in general? 
Some insight can be gained by considering the behaviour of arbitrary 
Green's functions at space-(imaginary) time points $z_1=\tau_1+ir_1$, $z_2=\tau_2+ir_2$, $\ldots$
Very close 
to the boundary we expect non-universal behaviour. 
If all points $r_i$ are very far from the boundary, $r_i\gg |z_j-z_k|$, 
then we expect to recover the bulk behaviour, unaffected by the boundary 
interactions. This behaviour is conformally invariant.   However, if 
the time-separations of some of the points are larger than or of 
order of the distances from the boundary, which are themselves 
large compared to microscopic scales, then the boundary 
still affects the Green's functions. We expect it to do 
so in a conformally invariant way.  We have a sort of 
conformally invariant termination of the bulk conformal 
behaviour, which is influenced by the universality class 
of the boundary, encoded in a conformally invariant boundary condition. 
Note that the RG flow being discussed here is entirely restricted 
to boundary interactions.  The bulk terms in the effective Hamiltonian 
do not renormalize, in our description; they sit at a bulk critical point. 
We do not expect finite range interactions, localized near $r=0$ to 
produce any renormalization of the bulk behaviour. All of the RG 
flows, which play such an important role in these lectures, are boundary 
RG flows. 

It actually turns out to be extremely important to go slightly beyond merely identifying 
the low energy fixed point, and to consider in more detail how it is approached, 
as the energy is lowered. As is usual in RG analyses, this is controlled 
by the leading irrelevant operator (LIO) at this fixed point. This is a boundary 
operator, defined in the theory with the conformally invariant boundary condition (CIBC) characteristic 
of the fixed point. It is important to realize that, in general, the set 
of boundary operators which exist depends on the CIBC.

 In the case at hand, 
the simplest version of the Kondo model, the boundary operators are simply 
constructed out of the fermion fields, which now obey the BC of Eq. (\ref{BCs}). 
It is crucial to realize that the impurity spin operator cannot appear 
in the low energy effective Hamiltonian because it is screened. Thus the 
LIO is constructed from fermion fields only. It must be SU(2) invariant. 
In general, the operator $\psi^{\dagger \alpha}(0)\psi_\alpha (0)$ might appear. 
This has dimension 1 and is thus marginal.  Note that 1 is the marginal 
dimension for boundary operators in a (1+1) dimensional CFT since 
these terms in the action are integrated over time only, not space. 
If we restrict outselves to the PH symmetric case, then this operator 
cannot appear since it is odd under the PH transformation, $\psi \to \psi^\dagger$. 
 Thus  we must turn to 4-fermion operators, of dimension 2, which are irrelevant. 
There are 2 operators allowed by SU(2) symmetry, $J(0)^2$ and 
$\vec J(0)^2$, where the charge and spin currents are defined as:
\be J\equiv \psi^{\dagger \alpha}\psi_{\alpha},\ \  \vec J\equiv \psi^{\dagger \alpha}
{\vec \sigma_\alpha^\beta \over 2}\psi_\beta .\ee
Here I am suppressing $L$, $R$ indices.  I work in the purely left-moving formalisim 
so all operators are left-movers. I will argue below that, since the Kondo interaction 
involves the spin degrees of freedom, the $\vec J^2$ term in the effective Hamiltonian 
has a much larger coefficient that does the $J^2$ term. More precisely, 
we expect that coefficient of the $\vec J^2$ term to be of order $1/T_K$. By power counting, 
 it must have a coefficient with dimensions of inverse energy. $1/T_K$ is the largest 
possible coefficient (corresponding to the lowest characteristic energy scale) 
that could occur and there is no reason why it should  not occur in general. 
Another way of looking at this is that the low energy effective Hamiltonian 
has a reduced cut-off of order $T_K$ (or $T_K/v_F$ in wave-vector units). 
The coefficient of $\vec J^2$ is of order the inverse cut-off. By contrast, 
I shall argue below that the coeffient of $J^2$ is of order 
$1/D_0\ll 1/T_K$.  Thus, this term can be ignored. The precise 
value of the coefficient of $\vec J^2$ is not known; but neither 
have I yet given a precise definition of $T_K$. Unfortunately, 
there are a large number of definitions of characteristic energy scales in use, 
referred to as $T_K$ among other things.  One possibility is 
to fix a definition of $T_K$ from the coupling constant of the LIO.  
\be H= {v_F\over 2\pi }i\int_{-\infty}^\infty dr \psi_L^\dagger {d\over dr}\psi_L
-{1\over 6T_K}\vec J_L(0)^2.\label{Heff}\ee
The factor of $1/6$ is inserted for convenience.  The fact that the 
sign is negative has physical significance and in principle can 
only be deduced from comparison to  other calculations (or experiments). 
Note that this Hamiltonian is defined with the low energy effective BC of 
Eq. (\ref{BCs}).  This means that the unfolding transformation 
used to write Eq. (\ref{Heff}) is changed by a minus sign.  i.e. Eq. (\ref{fold})
is modified to:
\be \psi_R(r)=-\psi_L(-r),\ \  (r>0).\ee
With this effective Hamiltonian in hand, we may calculate various 
physical quantities in perturbation theory in the LIO, i.e. perturbation 
theory in $1/T_K$. This will be discussed in detail later but 
basic features follow from power counting. An important result 
is for the impurity magnetic susceptibility. The susceptibility is:
\be \chi (T)\equiv {1\over T}<(S^z_T)^2>,\ee
where $\vec S_T$ is the {\it total} spin operator including 
both impurity and electron spin operators. The impurity 
susceptibility is defined, motivated by experiments, 
as the difference in susceptibilities of samples with 
and without the impurity. In practice, for a finite density of 
impurities, it is the term in the virial expansion of the susceptibility 
of first order in the impurity density $n_i$: 
\be \chi = \chi_0+n_i\chi_{imp}+\ldots \ee
Ignoring the LIO, $\chi_{imp}$ vanishes at low $T$. This 
follows because the Hamiltonian of Eq. (\ref{Heff}) is 
translationally invariant. A simple calculation, reviewed later, 
shows that, to first order in the LIO:
\be \chi_{imp}\to {1\over 4T_K}.\label{chil}\ee
This is the leading low $T$ result; corrected by a power 
series in $T/T_K$.  On the other hand, the high $T$ result at $T\gg T_K$, 
in the scaling limit of small $\lambda_0$, is:
\be \chi \to {1\over 4T},\label{chih}\ee
the result for a decoupled impurity spin. 
Our RG, BCFT methods only give the susceptibility 
in the low $T$ and high $T$ limits.  More powerful 
machinery is needed to also calculate it throughout the crossover, when $T$ 
is of order $T_K$. Such a calculate has been done accurately using the 
Bethe ansatz solution,\cite{Andrei,Weigmann} giving
\be \chi (T)={1\over 4T_K}f(T/T_K)\ee
where $f(x)$ is a universal scaling function.  The asymptotic 
results of Eqs. (\ref{chil}) and (\ref{chih}) are obtained. 
While the RG, BCFT methods are generally restricted to low energy 
and high energy limits (near the RG fixed points) they have the advantages 
of relative simplicity and generality (i.e. they are not restricted 
to integrable models). The impurity entropy (or equivalently the impurity 
specific heat) has similar behavior, with a contribution in first order 
in the LIO:
\be S_{imp}\to {\pi^2T\over 6T_K}.\ee
Again, $S_{imp}$ is a universal scaling function of $T/T_K$, approaching 
$\ln 2$, the result for a decoupled impurity, at high $T$. 
It is also possibly to calculate the impurity contribution to the 
electrical resistivity due to scattering off a dilute random array of impurities, 
using the Kubo formula. In this case, there is a contribution from the 
fixed point Hamiltonian itself, [i.e. from the modified BC of Eq. (\ref{BCs})]
even without including the LIO. This modified BC is equivalent to 
a $\pi /2$ phase shift in the s-wave channel (in the PH symmetric case). 
We may simply take over standard formulas for scattering from 
non-magnetic impurities, which make a contribution to the resistivity 
expressed entirely in terms of the phase shift at the Fermi energy. 
A phase shift of $\pi /2$ gives the maximum possibly resistivity, the
 so-called unitary limit:
\be \rho_u={3n_i\over (ev_F\nu )^2}.\label{rhoU}\ee
 The correction to the unitary limit 
can again be calculated in perturbation theory in the LIO.  In this case 
the leading correction is second order:
\be \rho (T)\approx \rho_u\left[1-{\pi^4T^2\over 16T_K^2}\right]\ \  (T\ll T_K).\label{rhofl}\ee
Again, at $T\gg T_K$ we can calculate $\rho (T)$ in perturbation theory 
in the Kondo interaction:
\be \rho (T)\approx \rho_u{3\pi^2\over 16}{1\over \ln^2(T/T_K)}\ \  (T\gg T_K).\ee
In between a scaling function of $T/T_K$ occurs.  It has so far not been 
possible to calculate this from the Bethe Ansatz but fairly accurate results 
have been obtained using Numerical Renormalization Group methods.  For a 
review see [\onlinecite{Bulla}].

This perturbation theory in the LIO is referred to as ``Nozi\`eres local Fermi liquid theory'' (FLT).
This name is highly appropriate due to the close parallels with Fermi liquid theory 
for bulk (screened) Coulomb interactions.  

\section{Multi-Channel Kondo Model}
\begin{figure}[th]
\begin{center}
\includegraphics[clip,width=8cm]{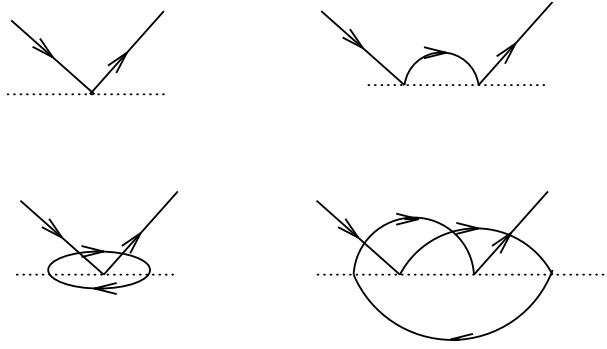}
\caption{Feynman diagrams contributing to renormalization of the
Kondo coupling constant to third order.}
\label{fig:pert}
\end{center}
\end{figure}

\begin{figure}
\epsfxsize=8 cm
\centerline{\epsffile{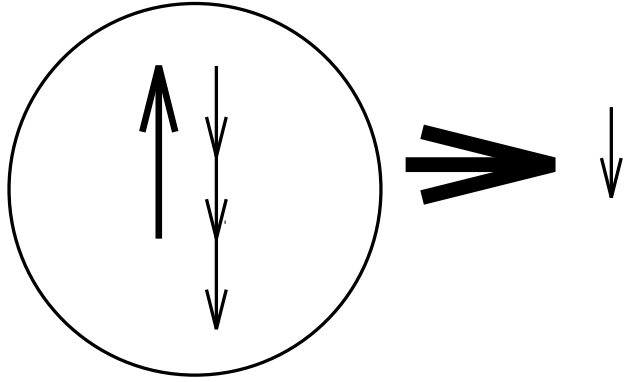}}
\caption{Formation of an effective spin at strong
Kondo coupling.  $k=3$, $s=1$ and
$s_{\hbox{eff}}=1/2$.}
\label{fig:seff}
\end{figure}

In this lecture, I will generalize the BCFT analysis of the simplest Kondo model to 
the multi-channel case.\cite{Nozieres3}  I will give a fairly sketchy overview of this subject here; 
more details are given in my previous summer school lecture notes [\onlinecite{Affleckrev}]. 
One now imagines $k$ identical ``channels'' all interacting 
with the same impurity spin, preserving an $SU(k)$ symmetry. In fact, it is 
notoriously difficult to find physical systems with such $SU(k)$ symmetry, so 
the model is an idealization. Much effort has gone into finding (or creating) 
systems realizing the $k=2$ case, with recent success. Jumping immediately 
to the  continuum limit, the analogue of Eq. (\ref{Hb}) 
\be H= {v_F\over 2\pi }i\int_0^\infty dr\left(\psi_L^{\dagger j} {d\over dr}\psi_{Lj}-\psi_R^{\dagger j}{
d\over dr}
\psi_{Rj}
\right)+v_F\lambda \psi_L^{\dagger j}(0){\vec \sigma \over 2}\psi_{Lj}(0)\cdot \vec S .
\label{Hmc}\ee
The repeated index $j$ is summed from $1$ to $k$; the spin indices are not written explicitly. 
The same ``bare'' BC, Eq. (\ref{freebc}) is used. The RG equations are only trivially 
modified:
\be {d\lambda\over d\ln D}=-\lambda^2 +{k\over 2}\lambda^3+\ldots \ee
The factor of $k$ in the $O(\lambda^3)$ term follows from the closed loop 
in the third diagram in Fig. (\ref{fig:pert}). We again conclude that a small bare coupling 
gets larger under the RG. 
However, for general choices of $k$ and 
the impurity spin magnitude, $S$, it can be readily seen that the 
simple strong-coupling BC of Eq. (\ref{BCs}) does not occur at the infrared fixed point. 
This follows from Fig. (\ref{fig:seff}). If the coupling flowed to infinity then 
we would expect $k$ electrons, one from each channel, to go into a symmetric 
state near the origin (at the first site in the limit of a strong bare coupling). 
They would form a total spin $k/2$. The antiferromagnetic coupling to the impurity 
of spin $S$ would lead to a ground state of size $|S-k/2|$. It is important 
to distinguish 3 cases: $S<k/2$, overscreened, $S>k/2$, underscreened and 
$S=k/2$ exactly screened. It turns out this strong coupling fixed point is 
stable in the underscreened and exactly screened cases only. In the exactly 
screened case this follows from the same considerations as for $S=1/2$, $k=1$, 
discussed in the previous lecture, which is the simplest exactly screened case. 
Otherwise, further considerations of this strong coupling fixed point are neccessary. 
This effective spin will itself have a Kondo coupling to the conduction 
electrons, obeying the strong coupling BC of Eq. (\ref{BCs}). This 
is clearest in the lattice model discussed in the previous lecture.  The 
effective spin is formed between  the impurity spin and the electrons on site $0$. 
If $J\gg t$, electrons from site $1$ can make virtual transitions onto site $0$, 
producing a high energy state. Treating these in second order gives an 
effective Kondo coupling to site $1$ with
\be J_{eff}\propto {t^2\over J}.\ee
The sign of $J_{eff}$ is clearly crucial; if it is ferromagnetic, 
then it would renormalize to zero.  In this case, the 
strong coupling fixed point is stable.  [Note that $J_{eff}\to 0$ 
corresponds to the original $J\to \infty$.] However, if it 
is antiferromagnetic then $J_{eff}$ gets larger as we lower the energy scale. 
This invalidates the assumption that the strong coupling fixed point is stable. 
[$J_{eff}$ getting larger corresponds to $J$ getting smaller.] It is 
not hard to see that $J_{eff}>0$ for the overscreened case and $J_{eff}<0$ 
for the underscreened case. [This calculation can be done in two steps. 
First, consider the sign of the exchange interaction between the electron 
spin on site $0$ and on site $1$. This is always antiferromagnetic, 
as in the Hubbard model.  Then consider the relative orientation 
of the electron spin on site $0$ and $\vec S_{eff}$.  These 
are parallel in the overscreened case but anti-parallel in the underscreened case.]
So, in the underscreened case, we may apply the results of the previous lecture 
with minor modifications. There is an impurity spin of size $S_{eff}$ in 
the low energy Hamiltonian but it completely decouples at the infrared fixed point. 
The overscreened case is much more interesting.  It is a ``non-Fermi liquid''
 (NFL). 

To solve this case I introduce the idea of a conformal embedding (CE). This is 
actually useful for various other BCFT problems. It is a generalization of 
the idea of bosonization, a powerful technique in (1+1) dimensions. 

We start by considering a left-moving single component (no channels, no spin) fermion field with
Hamiltonian density: \begin{equation} {\cal
H}=\frac{1}{2\pi}\psi_L^{\dagger}i\frac{d}{dx}\psi_L.
\end{equation}

Define the current (=density) operator, \be J_L(t+x)
=:\psi_L^\dagger\psi_L:(x,t) =\lim_{\epsilon\rightarrow  0}[\psi_L^\dagger (x)\psi_L
(x+\epsilon)-\langle 0|\psi_L^\dagger (x)\psi_L(x+\epsilon)|0\rangle ].
\ee
(Henceforth we generally drop the subscripts ``L'' and the time argument.  The double dots 
denote normal ordering: creation operators on the right.)
We will reformulate the theory in terms of currents (the key to
bosonization).  Consider: \begin{eqnarray}
J(x)J(x+\epsilon) &=& :\psi^\dagger (x)\psi (x)\psi^\dagger (x+\epsilon )
\psi (x+\epsilon ):+[:  \psi^\dagger
(x)\psi(x+\epsilon):+:\psi (x)\psi^\dagger
(x+\epsilon):]G(\epsilon)+ G(\epsilon)^2\nonumber\\ G(\epsilon)&=&
\langle 0|\psi(x)\psi^\dagger (x+\epsilon)|0\rangle
=\frac{1}{-i\epsilon} .\end{eqnarray} 
 By Fermi statistics the
4-Fermi term vanishes as $\epsilon\rightarrow 0$ \begin{equation}
:\psi^\dagger (x)\psi(x)\psi^\dagger (x)\psi(x):\  =-:\psi^\dagger
(x)\psi^\dagger (x)\psi(x)\psi(x):\ =0 .\end{equation} The second
term becomes a derivative,  \begin{eqnarray} \lim_{\epsilon
\rightarrow  0}[J(x)J(x+\epsilon )+\frac{1}{\epsilon^2}]
&=&\lim_{\epsilon\rightarrow  0}\frac{1}{-i\epsilon}[:\psi^\dagger
(x)\psi(x+\epsilon):- :\psi^\dagger
(x+\epsilon)~~\psi(x):]\nonumber\\ &=&2i:\psi^\dagger
\frac{d}{dx}\psi:\nonumber\\ {\cal H}&=&\frac{1}{4\pi}
J(x)^2+\mbox{constant} .\end{eqnarray} Now consider the commutator,
$[J(x),J(y)]$. The quartic and quadratic terms cancel. We must be
careful about the divergent  c-number part,  \begin{eqnarray}
[J(x),J(y)]&=&
-\frac{1}{(x-y-i\delta)^2}+\frac{1}{(x-y+i\delta)^2},\ \
(\delta\rightarrow  0^+)\nonumber\\ &=&\frac{d}{dx}\left[
\frac{1}{x-y-i\delta}-\frac{1}{x-y+i\delta}\right] \nonumber\\
&=&2\pi i \frac{d}{dx} \delta(x-y). \end{eqnarray}
Here $\delta$ is an ultraviolet cut-off.

Now consider the free massless boson theory with Hamiltonian
density (setting $v_F=1$): \begin{equation} {\cal
H}=\frac{1}{2}\left( \frac{\partial\varphi}{\partial
t}\right)^2+\frac{1}{2}\left( \frac{\partial\varphi}{\partial
x}\right)^2,
 ~~~~[\varphi(x), \frac{\partial}{\partial t}\varphi (y)]=i\delta(x-y)
\end{equation}

We can again decompose it into the left and  right-moving parts,
\begin{eqnarray} ({\partial_t}^2-{\partial_x}^2)\varphi&=&
(\partial_t+\partial_x)(\partial_t-\partial_x)\varphi \nonumber\\
\varphi(x,t)&=&\varphi_L(x+t)+\varphi_R(x-t)\nonumber\\
(\partial_t-\partial_x)\varphi_L&\equiv&\partial_-\varphi_L=0,
{}~~\partial_+\varphi_R=0\nonumber\\
H&=&\frac{1}{4}(\partial_-\varphi)^2+\frac{1}{4}(\partial_+\varphi)^2=
\frac{1}{4}(\partial_-\varphi_R)^2+\frac{1}{4}(\partial_+\varphi_L)^2
\end{eqnarray}
where
\be \partial_{\pm}\equiv \partial_t\pm \partial_x.\ee
Consider the Hamiltonian density for a left-moving boson field:
\begin{eqnarray} {\cal H}&=&\frac{1}{4}(\partial_+\varphi_L)^2
\nonumber\\ {[}\partial_+\varphi_L(x), \partial_+\varphi_L(y)]&=&
[\dot\varphi+\varphi',\dot\varphi+\varphi']=2i\frac{d}{dx}\delta(x-y)
\end{eqnarray}
Comparing to the Fermionic case, we see that: \begin{equation}
J_L=\sqrt{\pi}\partial_+\varphi_L=\sqrt{\pi}\partial_+\varphi,
\end{equation} since the commutation relations and Hamiltonian are
the  same. That means the operators are the same  with  appropriate
boundary conditions.   

This equivalence becomes especially powerful when the fermions 
have several components. Consider the case at hand with 
2 spin components and $k$ channels. Clearly we can write the 
free fermion Hamiltonian in this case as:
\be {\cal H}(x)={1\over 4\pi}\lim_{\epsilon \to 0}\sum_{\alpha j}
:\psi^{\dagger \alpha j}\psi_{\alpha j}:(x)\ :\psi^{\dagger \alpha j}\psi_{\alpha j}:(x+\epsilon )\ 
+\hbox{constant}.\ee
It turns out to be very useful to use simple algebraic identifies to rewrite this in terms of 
charge, spin and channel current operators:
\bea J(x)&\equiv & :\psi^{\dagger \alpha i}\psi_{\alpha i}:\nonumber \\
\vec J &\equiv & \psi^{\dagger \alpha i}{\vec \sigma_{\alpha}^{\beta} \over 2}\psi_{\beta i}\nonumber \\
J^A&\equiv &\psi^{\dagger \alpha i}\left(T^A\right)^j_i\psi_{\alpha i}.\eea
Here the $T^A$'s are generators of $SU(k)$, i.e. a set of traceless Hermitian matrices obeying 
the orthonormality condition:
\be \hbox{Tr}T^AT^B={1\over 2}\delta^{AB}\ee
and hence the completeness relation:
\be \sum_A\left( T^A\right)^b_a\left(T^A\right)_c^d={1\over 2}\left[\delta^b_c\delta^d_a-{1\over k}
\delta^b_a\delta^d_c\right].\ee
They obey the commutation relations:
\be [T^A,T^B]=i\sum_Cf^{ABC}T^C,\ee
where the numbers $f^{ABC}$ are the structure constants of $SU(k)$. 
In the $k=2$ case, we may choose:
\be T^a\to {\sigma^a\over 2}.\ee
It is now straight forward to prove the following identity:
\be {\cal H} = {1\over 8\pi k}J^2+{1\over 2\pi
(k+2)}\vec J^2+{1\over 2\pi (k+2)}J^AJ^A.\label{Sug}\ee
The cofficients of each term are chosen so that the normal ordered products 
$:\psi^{\dagger \alpha a}\psi_{\alpha a}\psi^{\dagger \beta b}\psi_{\beta b}:$ and 
$:\psi^{\dagger \alpha a}\psi_{\alpha b}\psi^{\dagger \beta b}\psi_{\beta a}:$
 have zero 
coefficients. The current operators now obey the current algebras:
\bea [J(x),J(y)]&=&4\pi ik\delta{'} (x-y)\nonumber \\
{[}J^a(x),J^b(y)]&=&2\pi i\delta (x-y)\epsilon^{abc}J^c+\pi ik\delta^{ab}\delta^{'}(x-y)\nonumber \\
{[}J^A(x),J^B(y)]&=&2\pi i\delta (x-y)f^{ABC}J^C+2i\pi \delta^{AB}\delta^{'}(x-y).\label{com}\eea
The currents of different types (charge, spin, flavour) commute with each other. 
Thus the Hamiltonian is a sum of three commuting terms, for charge, spin and flavour, 
each of which is quadratic in currents and each of which is fully characterized 
by the current commutation relations. Upon including the right-moving degrees of freedom, 
we may define three independent field theories, 
for spin, charge and flavour with the corresponding Hamiltonians. The charge 
Hamiltonian is simply a free boson, with:
\be J=\sqrt{2\pi k}\partial_+\varphi .\ee
The spin and channel Hamiltonians are Wess-Zumion-Witten (WZW) non-linear $\sigma$-models 
(NL$\sigma$M)\cite{Witten,Knizhnik}
labelled $SU(2)_k$ and $SU(k)_2$. 
These can be written in terms of $SU(2)$ and $SU(k)$ bosonic matrix fields $g^\alpha_\beta (t,x)$ and $h^i_j(t,x)$ respectively. 
(These fields are Lorentz scalars; i.e. they have zero conformal spin.)  The corresponding current 
operators can be written in a form quadratic in these matrix fields. 
In the particular case, $k=1$, 
the corresponding $SU(2)_1$ WZW is simply equivalent to a free boson. The connection between 
the multi-component free fermion model and the sum of spin, charge and channel bosonic models 
is an example of a conformal embedding. All conformal towers in the free fermion finite size spectrum (FSS), 
with various BC's can be written as sums of products of conformal towers from the 3 constitutent models. 
Likewise, each local operator in the free fermion model is equivalent to a product of 
charge, spin and flavour local operators in the bosonic models. [Actually, this statement 
needs a little qualification.  It is literally true for free fermion operators which 
contain even numbers of fermion fields and have zero conformal spin. It is only true for the fermion fields themselves 
if we are allowed to define chiral components of the WZW matrix fields.]

We now adopt the purely left-moving representation of the Kondo model,  in Eq. (\ref{HL}). 
This has the advantage that the Kondo interaction can be written in terms of the spin current 
operators at the origin only. Thus, remarkably, the Kondo interaction is entirely 
in the spin sector:
\be H={1\over 2\pi (k+2)}\int_{-\infty}^\infty dr \vec J(r)^2+\lambda \vec J(0)\cdot \vec S+\ldots .
\label{Hsep}\ee
Here the $\ldots$ represents the charge and channel parts of the Hamiltonian which are non-interacting, 
decoupled from the impurity. 
An immediate consequence of this spin-charge-channel separated 
form of the Kondo Hamiltonian is that the Kondo interaction only appears in the spin sector. 
It is then reasonable to expect that the LIO at the Kondo fixed involves the spin operators only, 
corresponding to Eq. (\ref{Heff}).  
If we took Eq. (\ref{Hsep}) at face value this would appear to 
be exactly true. In fact, since Eq. (\ref{Hsep}) is only a low energy effective Hamiltonian, 
we can generate other operators, in the charge (and channel) sectors during intermediate 
stages of the RG. However, we expect all such operators to have much smaller coefficients, 
with the scale set by $D_0$ rather than $T_K$. This is the reason we ignored the 
irrelevant operator $J^2$ in the previous lecture. The marginal operator which could 
be added to the effective Hamiltonian when particle-hole symmetry is broken, is 
now seen to be purely a charge operator: $J_L=\sqrt{2\pi k}\partial_+\phi$.  Because 
it is linear in the charge boson its effects are easy to include and it is strictly marginal. 
Assuming it is small, we can simply ignore it.  It leads to a line of fixed points. 

It is interesting to observe that this Hamiltonian can be formally 
diagonalized, for a special value of $\lambda_c = 2/(k+2)$ by redefining the spin currents:
\be \tilde J^a(r)\equiv J^a(r)+2\pi \delta (r)S^a.\label{JS}\ee
It can readily be checked that the $\tilde J^a(r)$ obey the same commutation relations as 
in Eq. (\ref{com}).  Furthermore, for $\lambda = \lambda_c$, the interacting Hamiltonian 
reduces to:
\be H={1\over 2\pi (k+2)}\int_{-\infty}^\infty dr [\tilde J^a(r)]^2+\ldots \ee
This suggests that there might be an infrared stable fixed point of the RG 
at an intermediate value of $\lambda$ corresponding to a BCFT. To make this idea 
more quantitative, we must use the full apparatus of BCFT. See J. Cardy's lecture 
notes from this Summer School for a review of this subject.\cite{CardyH} 

A central idea of Cardy's BCFT is that one should represent CI BC's by {\it boundary states}.\cite{Cardy1}
These states contain all low energy information about a BCFT.  From them one 
can construct the finite size spectrum (with any pair of CI BC's at the two 
ends of a finite system), OPE coefficients, boundary operator content, and 
all other universal properties. To each CFT there is a set of possible 
boundary states (i.e. a set of possible CIBC's). In general, a complete 
classification of {\it all} (conformally invariant) boundary states is not available. 
However, in the case of rational CFT's, with a finite number of conformal 
towers, a complete classification is available. The boundary states are 
in one-to-one correspondance with the conformal towers, i.e. with the 
primary operators. One can obtain the complete set of boundary states, 
from a reference state by a process of fusion with primary operators. 
Our strategy for finding the low energy fixed point of the general Kondo models 
(and various other quantum impurity probems) is to first identify the 
boundary state corresponding to the trivial boundary conditions of Eq. (\ref{freebc}). 
We then obtain the CIBC corresponding to the low energy fixed point by 
fusion with an appropriate primary operator. The choice of primary 
operator is inspired by the mapping in Eq. (\ref{JS}).

The conformal towers of the $SU(2)_k$ WZW model are labelled by the spin 
of the ``highest weight state'' (i.e. the lowest energy state).\cite{Knizhnik,Gepner,Francesco} There is 
one conformal tower for each spin $j$ with:
\be j=0,1/2,1,\ldots k/2.\ee
These primary fields have zero conformal spin and left and right scaling dimensions:
\be \Delta = {j(j+1)\over k+2}.\label{dim}\ee
We may associate them with the spin part of the fermion operators $(\psi_L)^{n}$. The 
largest possible spin we can get this way, from $n=k$, anti-symmetrized with respect to flavour, 
is $j=k/2$. 
The fusion rules are:
\begin{equation} j \otimes j' =
|j-j'|, |j-j'|+1, |j-j'|+2, \ldots ,\hbox{min} \{ j+j', k-j-j'\}
.\label{frules}\end{equation}   Note that this generalizes the ordinary angular
momentum addition rules
 in a way which is consistent with the conformal tower structure of
the theories (i.e. the fact that primaries only exist with $j\leq
k/2$).

Based on Eq. (\ref{JS}), we expect that the infrared stable fixed point 
of the $k$-channel Kondo model with a spin $S$ impurity corresponds to 
fusion with the spin $j=S$ primary operator, whenever $S\leq k/2$. 
Note that the spin quantum numbers of the conformal towers match 
nicely with the over/under screening paradigm. For $S>k/2$ we obtain the 
infrared stable fixed point by fusion with the maximal spin primary 
operator of spin $k/2$. Thus the boundary state at the infrared (``Kondo'') 
fixed point is related to the free fermion boundary state by:\cite{CardyH,Cardy1}
\be <j0|\hbox{Kondo}>=<j0|\hbox{free}>{S^j_S\over S^j_0}.\label{Bstate}\ee
Here the boundary states are expanded in the Ishibashi states corresponding 
to the Kac-Moody conformal towers of spin $j$ and $|j0>$ labels 
the ground state of the spin $j$ conformal tower. $S^j_{j'}$ is the 
modular S-matrix\cite{Kac} for $SU(2)_k$:
\be  S^j_{j'} (k) =
\sqrt{2\over 2+k}\sin \left[{\pi (2j+1)(2j'+1)\over
2+k}\right], 
\label{Smatrix}\ee
This ``fusion rule hypothesis'' leads immediately to various predictions 
about the low energy behaviour which can be tested against numerical 
simulations, Bethe Ansatz calculations and experiments. 
One important comparison involves the finite size spectrum. With 
the free BC's of Eq. (\ref{freebc}) the FSS can be written 
as a sum of direct products of conformal towers from spin, charge 
and channel sectors, $(Q,j,j_c)$.  Here $Q$ is the charge 
of the highest weight state (measured from the charge of the ground state) 
and $j_c$ is a shorthand notation 
for the $SU(k)$ quantum numbers of the highest weight state 
of the $SU(k)_2$ WZW model for the channel degrees of freedom.  
In the important example $k=2$ it corresponds literally to a 
second set of $SU(2)$ ``pseudo-spin'' quantum numbers. To obtain 
the spectrum at the infrared fixed point, one replaces the 
spin-$j$ conformal tower by a set of spin conformal towers using 
the $SU(2)_k$ fusion rules of Eq. (\ref{frules}) with $j'$ replaced by $S$, 
the impurity spin magnitude, in the over and exactly screened cases.
 (In the underscreened case, $S$ should be replaced by $k/2$.) Since the 
full spectrum of each conformal tower is easily constructed, the 
``fusion rule hypothesis'' predicts an infinite number of finite size 
energy levels.  These can be compared to the results of Numerical Renormalization 
Group (NRG) calculations. These calculations give the spectrum of a finite 
chain of length $l$, with the impurity spin at one end, like the 
tight-binding model of Eq. (\ref{lattice}). These spectra reveal 
an interesting cross-over behaviour. For a weak Kondo coupling 
and a relatively short chain length the spectrum is essentially 
that of the zero Kondo coupling model: i.e. the conformal spectrum 
with the BC of Eq. (\ref{freebc}) factored with the decoupled impurity spin. 
However, as the chain length increases this spectrum shifts.  The characteristic 
cross-over length is 
\be \xi_K\equiv v_F/T_K\propto \exp [1/\lambda_0].\ee
For longer chain lengths the FSS predicted by the our BCFT methods is 
observed, for the low energy part of the spectrum. (In principle, the 
smaller the bare Kondo coupling and the longer the chain length 
the more states in this conformal BCFT spectrum are observed.) 
In [\onlinecite{Affleck5}], for example, the first 6 energy levels (most of which 
are multiply degenerate) were compared, for the $k=2$ case, obtaining excellent agreement.

\subsection{Impurity Entropy} We define the impurity entropy as:
\begin{equation} S_{\hbox{imp}}(T) \equiv \lim_{l\to
\infty}[S(l,T)-S_0(l,T)],\label{Simpdef}\end{equation} where
$S_0(l,T)$ is the free fermion entropy, proportional to $l$, in the
absence of the impurity.
 Note that, for zero Kondo coupling,
$S_{\hbox{imp}}=\ln [s(s+1)],$ simply reflecting the groundstate
degeneracy of the free spin.  In the case of exact screening,
($k=2s$), $S_{\hbox{imp}}(0)=0$.  For underscreening,
\begin{equation} S_{\hbox{imp}}(0) = \ln [s'(s'+1)],\end{equation}
where $s'\equiv s-k/2$.  What happens for overscreening?
Surprisingly, we will obtain, in general, the log of a non-integer,
implying a sort of ``non-integer groundstate degeneracy''.

To proceed, we show how to calculate $S_{\hbox{imp}}(0)$ from the
boundary state.  All calculations are done in the scaling limit,
ignoring irrelevant operators, so that  $S_{\hbox{imp}}(T)$ is a
constant, independent of $T$, and characterizing the particular
boundary condition.  It is important, however, that we take the
limit $l\to \infty$ first, as specified in Eq. (\ref{Simpdef}), at
fixed, non-zero $T$.  i.e. we are interested in the limit, $l/\beta
\to \infty$.  Thus it is convenient to use the  expression for
the partition function,\cite{CardyH,Cardy1} $Z_{AB}$:
\begin{equation} Z_{AB} = \sum_a<A|a0><a0|B>\chi_a(e^{-4\pi
l/\beta})\to e^{\pi lc/6\beta}<A|00><00|B>.\end{equation}  Here
$|a0>$ labels the groundstate in the conformal tower of the
operator $O_a$ and $\chi_a$ is the corresponding character.  $c$ is the conformal anomaly.   Thus the free
energy is: \begin{equation} F_{AB} = -\pi cT^2l/6-T\ln
<A|00><00|B>.\end{equation} The first term gives the specific heat:
\begin{equation} C=\pi cTl/3\end{equation} and the second gives the
impurity entropy: \begin{equation} S_{\hbox{imp}} = \ln
<A|00><00|B>.\end{equation}  This is a sum of contributions from
the two boundaries,  \begin{equation}
S_{\hbox{imp}}=S_A+S_B.\end{equation}  Thus we see that the
``groundstate degeneracy'' $g_A$, associated with boundary
condition A is: \begin{equation} \exp [S_{\hbox{imp}A}] =
<A|00>\equiv g_A.\end{equation}  Here we have used our freedom to
choose the phase of the boundary state so that $g_A>0$.  For our
original, anti-periodic, boundary condition, $g=1$. For the Kondo
problem we expect the low T impurity entropy to be given by the
value at the infrared fixed point.  Since this is obtained by
fusion with the spin-s (or k/2) operator, we obtain from Eq.
(\ref{Smatrix}), \begin{equation} g = {S^0_s\over
S^0_0}={\sin [\pi
(2s+1)/(2+k)]\over \sin [\pi /(2+k)]}.\end{equation} 
 This formula agrees exactly with the
Bethe ansatz  result\cite{Tsvelik} and has various
interesting properties.  Recall that in the case of exact or
underscreening ($s\geq k/2$) we must replace $s$ by $k/2$ in this
formula, in which case it reduces to 1.  Thus the groundstate
degeneracy is 1 for exact screening.  For underscreening we must
multiply $g$ by $(2s'+1)$ to account for the decoupled, partially
screened impurity.  Note that, in the overscreened case, where $s<
k/2$, we have: \begin{equation} {1\over 2+k}<{2s+1\over
2+k}<1-{1\over 2+k},\end{equation} so $g>1$. In the case $k\to
\infty$ with s held fixed, $g\to 2s+1$, i.e. the entropy of the
impurity spin is hardly reduced at all by the Kondo interaction,
corresponding to the fact that the critical point occurs at weak
coupling.  In general, for underscreening: \begin{equation}
1<g<2s+1.\end{equation} i.e. the free spin entropy is somewhat
reduced, but not completely eliminated.  Furthermore, $g$ is not,
in general, an integer. For instance, for $k=2$ and $s=1/2$,
$g=\sqrt{2}$. Thus we may say that there is a non-integer
``groundstate degeneracy''.  Note that in all cases the groundstate
degeneracy is reduced under renormalization from the zero Kondo
coupling fixed point to the infrared stable fixed point.  This is a
special case of  a general result: {\it the
groundstate degeneracy always decreases under renormalization.}
This is related to Zamolodchikov's
c-theorem\cite{Zamolodchikov2} which states that the conformal
anomaly parameter, c, always decreases under renormalization.  The
intuitive explanation of the c-theorem is that, as we probe lower
energy scales, degrees of freedom which appeared approximately
massless start to exhibit a mass.  This freezes out their
contribution to the specific heat, the slope of which can be taken
as the definition of c.  In the case of the ``g-theorem'' the
intuitive explanation is that, as we probe lower energy scales,
approximately degenerate levels of impurities exhibit small
splittings, reducing the degeneracy.

A ``perturbative'' proof of the g-theorem was 
given in [\onlinecite{Affleck6}] where RG 
flow between two ``nearby'' boundary RG fixed points with 
almost the same values of $g$ was considered. A general 
proof was given in [\onlinecite{Friedan}]. 

\subsection{Resistivity/Conductance}
In this subsection I consider the resistivity, due to scattering from a dilute array of $k$-channel Kondo 
impurities,\cite{Affleck6} 
and the closely related conductance through a single $k$-channel impurity. This latter quantity, in 
the $k=2$ case, was recently measured in quantum dot experiments, as I discuss in the next lecture. 
Using the Kubo formula, these quantities can be expressed in terms of the single-electron 
Green's function.  Due to the $\delta$-function nature of the Kondo interaction, the 
exact retarded Green's function (in 1, 2 or 3 dimensions) with a single impurity at $r=0$ can be written as:
\be G(\vec r, \vec r';\omega )=G_0(|\vec r-\vec r'|,\omega )+G_0(r,\omega ){\cal T}(\omega )G_0(r',\omega ).
\label{Tmatrix}\ee
Here $G_0$ is the non-interacting Green's function.  The function ${\cal T}$, which depends on the 
frequency only, not the spatial co-ordinates, is known as the ${\cal T}$-matrix. Note 
that I am using a mixed space-frequency representation of the Green's function, which is 
invariant under time-translations, but not space-translations. The only thing which 
distinguishes the dimensionality of space is $G_0$. 

In the case of a dilute random array of impurities, in D=3, the Green's function, 
to first order in the impurity concentration, $n_i$, can be written exactly as:
\be G(|\vec r -\vec r'|,\omega )={1\over G_0^{-1}(\vec r-\vec r'|,\omega )-\Sigma (\omega )},\ee
where the self-energy is given by:
\be \Sigma (\omega ) = n_i{\cal T}(\omega ).\ee
(Translational invariance is restored after averaging over impurity positions.) 
The single-electron life-time is given by:
\be \tau^{-1}(\omega )=\hbox{Im}\Sigma (\omega )\ee
and the finite temperature resistivity, $\rho (T)$, can be expressed in terms of this life-time 
by the standard formula:
\begin{equation}
{1\over \rho (T)} = {2e^2k\over 3m^2}\int{d^3p\over (2\pi )^3}\left[{-dn_F\over
d\epsilon_p}\right]\vec p^2\tau (\epsilon_p)\end{equation} 
where $n_F$ is the Fermi distribution function:
\be n_F\equiv {1\over \exp [\epsilon_p/T]+1}.\ee
At low temperatures, this integral is dominated by low energies so our 
field theory results can be used. A similar calculation, reviewed in the next lecture, 
expresses also the conductance through a single impurity in terms of  Im ${\cal T}$.

Thus, our task is to calculate the electron Green's function in the low energy 
1-dimensional effective field theory. In the zero temperature limit we 
may simply evaluate it at the Kondo fixed point.  At low finite temperatures 
we consider the correction from the LIO. At the fixed point, the chiral 
Green's functions, $<\psi_L^\dagger (r+i\tau )\psi_L(r'+i\tau ')>$,
$<\psi_R^\dagger (r-i\tau )\psi_R(r'-i\tau ')>$ are 
unaffected by the Kondo interaction.  We only need to consider $<\psi_L^\dagger (r,\tau )
\psi_R(r',\tau ')>$. By general methods of BCFT this behaves as a two-point function 
of left-movers with the right-mover reflected to the 
negative axis, $(-r',\tau ')$:
\be <0|\psi^{\dagger i\alpha}_L(\tau ,r)\psi_{Rj\beta}(r',\tau ')|0>={S_{(1)}
\delta^\alpha_\beta \delta^i_j\over (\tau -\tau ')+i(r+r')}.\ee
Only the constant, $S_{(1)}$, depends on the particular CIBC. For instance, if 
the BC is of free fermion type, $\psi_R(0)=e^{i\delta}\psi_L(0)$, then $S_{(1)}=e^{i\delta}$. 
In general, $S_{(1)}$ can be expressed in terms of the boundary state.\cite{Cardy3} Since the fermion 
field has spin $j=1/2$, the general expression is:
\be S_{(1)}={<1/2,0|A>\over <00|A>}.\ee
By comparing to the free fermion BC where $S_{(1)}=1$ and obtaining the Kondo BC by fusion,
 using Eq. (\ref{Bstate}), it follow that $S_{(1)}$ is given in terms of the modular $S$-matrix:
\be S_{(1)}={S^{1/2}_SS^0_S\over S^{1/2}_0S^0_S}={\cos \left[\pi (2S+1)/(2+k)\right]\over \cos \left[\pi
/(2+k)\right]}.\label{S1}\ee
The zero temperature ${\cal T}$-matrix can be expressed directly in terms of $S_{(1)}$:
\be {\cal T}(\omega )={-i\over 2\pi \nu}[1-S_{(1)}].\label{T0}\ee
In this limit, ${\cal T}$ is independent of $\omega$ and purely imaginary. This result 
follows from the definition, Eq. (\ref{Tmatrix}) of the ${\cal T}$-matrix upon using 
the free Green's function:
\be G_0(r,\omega_n)=2\pi ie^{\omega_nr}\left[\theta (-\omega_n)\theta (r)-\theta (\omega_n)\theta (-r)\right]\ee
where $\theta (x)$ is the step function together with the analytic continuation to real frequency:
\be \theta (\omega_n)\to \theta (\delta -i\omega )=1.\ee
In the exactly or underscreened case where we set $S=k/2$, Eq. (\ref{S1}) gives $S_{(1)}=-1$, 
the free fermion result with a $\pi /2$ phase shift.  This is the unitary limit 
resisitivity of Eq. (\ref{rhoU}), which I now define divided by a factor of $k$ since we have $k$ parallel channels. 
In general, at the non-Fermi liquid fixed points,
\be \rho (0)=\rho_U\left[{1-S_{(1)}\over 2}\right] \leq \rho_U.\ee

To calculate the leading corrections at low temperature (or frequency) we must 
do perturbation theory in the LIO. The LIO must be a boundary operator which exists 
under the CIBC's characterizing the fixed point and which, furthermore, 
respects all symmetries of the Hamiltonian. The set of boundary operators (for {\it any} 
CIBC) is a subset of the set of chiral operators in the bulk theory. This follows 
from the ``method of images'' approach to BCFT which expresses any local operator 
with left and right moving factors as a bilocal product of left-movers.  In the 
limit where the operator is taken to the boundary, we may use the OPE to express 
it in terms of local left-moving operators. The set of boundary operators which 
actually exist, for a given CIBC is in one-to-one correspondance with the set 
of conformal towers in the finite size spectrum with the corresponding boundary 
condition imposed {\it at both ends} of a finite system. This can be obtained 
by ``double fusion'' from the operator content with free fermion BC's. 
The  boundary operators with free BC's all have integer dimensions and include
Kac-Moody descendants of the identity operator, such as the current operators. 
Double fusion, starting with the identity operators 
corresponds to applying Eq. (\ref{frules}) {\it twice} starting 
with $j=0$ and $j'=S$. This gives operators of spin $j=0$, $1$, $\ldots$
min$\{2S,k-2S\}$. While this only gives back the identity operator $j=0$ 
for the exact or underscreened case, where $S=k/2$ it always gives $j=1$ 
(and generally higher integer spins) for the overscreened case.  We see from 
the dimensions, Eq. (\ref{dim}), that the spin-1 primary is the lowest 
dimension one that occurs with dimension
\be \Delta = {2\over 2+k}.\label{dimLIE}\ee
 None of these non-trivial primary operators 
can appear directly in the effective Hamiltonian since they are not rotationally 
invariant, having non-zero spin. However, we may construct descendant operators 
of spin 0. The lowest dimension spin zero boundary operator for all overscreened cases is
$\vec J_{-1}\cdot \vec \varphi$ where $\vec \varphi$ is the spin-1 primary operator. 
This is a first descendent, with scaling dimension $1+\Delta$. 
This is $<2$, the dimension of the Fermi liquid operator, $\vec J^2$ which can also occur.  
This is the LIO in the exact and underscreened cases, Eq. (\ref{Heff}).
Thus the effective Hamiltonian, in the overscreened case, may be written:
\be H=H_0-{1\over T_K^{\Delta}}\vec J_{-1}\cdot \vec \varphi .\label{Hnfl}\ee
Here $H_0$ is the WZW Hamiltonian with the appropriate BC. 
As usual, we assume that the dimensionful coupling constant multiplying the LIO 
has its scale set by $T_K$, the crossover scale determined by the weak coupling RG. 
We may take Eq. (\ref{Hnfl}) as our precise definition of $T_K$ (with the operator 
normalized conventionally). As in the Fermi liquid case, many different physical 
quantities can be calculated in lowest order perturbation theory in the LIO 
giving various generalized ``Wilson ratios'' in which $T_K$ cancels. 
One of the most interesting of these perturbative calculations is for the 
single-fermion Green's function, giving the 
${\cal T}$-matrix.  In the Fermi liquid case, the first order perturbation 
theory in the LIO gives a correction to the ${\cal T}$-matrix which is purely real. 
Only in second order do we get a correction to Im ${\cal T}$, leading 
to the correction to the resistivity of O$(1/T_K^2)$ in Eq. (\ref{rhofl}).  
On the other hand, a detailed calculation shows that first order 
perturbation theory in the non-Fermi liquid LIO of Eq. (\ref{Hnfl}), gives 
a correction to the ${\cal T}$-matrix with both real {\it and} imaginary parts 
and hence a correction to the resistivity of the form:
\be \rho (T)=\rho_U\left[{1-S_{(1)}\over 2}\right]\left[ 1-\alpha \left({T\over T_K}\right)^\Delta \right].
\label{rhonfl2}\ee
Here $\alpha$ is a constant which was obtained explicitly from the detailed perturbative calculation, 
having the value $\alpha = 4\sqrt{\pi}$ for the 2-channel $S=1/2$ case (for which $S_{(1)}=0$). Also note that 
the {\it sign} of the coupling constant in Eq. (\ref{Hnfl}) is {\it not} determined a priori. 
If we assumed the opposite sign, the $T$-dependent term in the resisitivy, Eq. (\ref{rhonfl2}) 
would switch. It is reasonable to expect this negative sign, for a weak bare coupling, 
since the resistivity is also a decreasing function of $T$ at $T\gg T_K$ where it 
can be calculated perturbatively in the Kondo coupling. An assumption of monotonicity of 
$\rho (T)$ leads 
to the negative sign in Eq. (\ref{Hnfl}). In fact, this negative sign has 
recently been confirmed by experiments, as I will discuss in the next lecture. 

A number of other low energy properties of the non-Fermi liquid Kondo fixed points have 
been calculated by these methods, including the $T$-dependence of the entropy and susceptibility 
and space and time dependent Green's function of the spin density, but I will not take the time to review them here.

\section{Quantum Dots: experimental realizations of one and two channel Kondo models}
In this lecture I will discuss theory and experiments on quantum dots, as experimental realizations 
of both single and two channel Kondo models. 
\subsection{Introduction to quantum dots}
Experiments on gated semi-conductor quantum dots begin with 2 dimensional electron gases (2DEG's) in 
semi-conductor heterostructures, usually GaAs-AlGaAs.  (These are the same types of semi-conductor
 wafers used for quantum Hall effect experiments.) A low areal density of electrons is trapped 
in an inversion layer between the two different bulk semiconductors. Great effort goes into 
making these 2DEG's very clean, with long scattering lengths. Because the electron density 
is so low compared to the inter-atomic distance, the dispersion relation near the Fermi energy 
is almost perfectly 
quadratic with an effective mass much lower than the free electron's. The inversion layer 
is located quite close to the upper surface of the wafer (typically around 100 nm. below it).  Leads 
are attached to the edges of the 2DEG to allow conductance measurements. In addition several leads 
are attached to the upper surface of the wafer, to apply gate voltages to the 2DEG, which can vary 
on distances of order .1 microns. Various types of quantum dot structures can be built on the 2DEG 
using the gates. A simple example is a single quantum dot, a roughly circular puddle of electrons, 
with a diameter of around .1 $\mu$.  The quantum dot is separated from the left and right 
regions of the 2DEG by large electroscatic barriers so that there is a relatively small 
rate for electrons to tunnel from the dot to the left and right regions of the 2DEG. In simple 
devices, the only appreciable tunnelling path from electrons from left to right 2DEG regions 
is through the quantum dot.  Because the electron transport in the 
2DEG is essentially ballistic, the current is proportional to the voltage 
difference $V_{sd}$ (source-drain voltage) between the leads, for small $V_{sd}$, rather than the electric field. The linear 
conductance, 
\be I=GV_{sd}\ee
(and also non-linear conductance) is measured versus $T$ and the various gate voltages. 

An even simpler device of this type does not have a quantum dot, but just a single point 
contact between the two leads. (The quantum dot devices have essentially two point 
contacts, from left side to dot and from dot to right side.) As the barrier height 
of the point contact is raised, so that it is nearly pinched off, it is found that 
the conductance, at sufficiently low $T$, has sharp plateaus and steps with the conductance 
on the plateaus being $2ne^2/h$, for integer $n$. $2e^2/h$ is the conductance of an 
ideal non-interacting one-dimensional wire, with the factor of $2$ arising from electron spin. 
This can be seen from a Landauer approach. Imagine left and right reservoirs at different 
chemical potentials, $\mu_R$ and $\mu_L-eV_{sd}$, with each reservoir emitting electrons 
to left and right into wires with equilibrium distributions characterized by 
different chemical potentials:
\bea I&=&-2e\int_0^\infty {dk\over 2\pi}v(k)[n_F(\epsilon_k-\mu +eV_{sd})-n_F(\epsilon_k-\mu )]\nonumber \\
G&=&2e^2\int_0^\infty {d\epsilon \over 2\pi}{dn_F\over d\mu }(\epsilon_k-\mu )={2e^2\over h}n_F(-\mu ),
\eea
where I have inserted a factor of $\hbar$, previously set equal to one, in the last step. Thus, 
provided that $\mu \gg k_BT$, $G=2e^2/h$ for an ideal one-dimensional conductor.  A wider 
non-interacting wire would have $n$ partially occupied bands and a conductance of $2ne^2/h$. 
As the point contact is progressively pinched off, it is modelled as a progressively narrower 
quantum wire with fewer channels, thus explaining the plateaus. Because the gate voltage varies 
gradually in the 2DEG, backscattering at the constriction is ignored; otherwise the 
conductance of a single channel would be $2e^2T_r/h$ where $T_r$ is the transmission probability. 

The tunnel barriers separating the quantum dot from the left and right 2DEG regions are 
modeled as single channel point contacts. Nonetheless, as the temperature is lowered 
the conductance through a quantum dot often tends towards zero. This is associated 
with the Coulomb interactions between the electrons in the quantum dot. Although it 
may be permissable to ignore Coulomb interactions in the leads this is not 
permissable in the quantum dot itself. A simple and standard approach is 
to add a term to the Hamiltonian of the form $Q^2/(2C)$ where $Q$ is the charge 
on the quantum dot and $C$ is its capacitance. In addition, a gate voltage, $V$, is 
applied to the dot, so that the total dot Hamiltonian may be written:
\be H_d={U\over 2}(\hat n-n_0)^2,\ee
where $\hat n$ is the number operator for electrons on the dot
 and $n_0\propto V$. 
An important dimensionless parameter is $t/U$ where $t$ is  the tunnelling 
amplitude between leads and dot.  If $t/U\ll 1$ then the charge on 
the dot is quite well-defined and will generally stay close to $n_0$ 
with virtual fluctuations into higher energy states with $n=n_0\pm 1$. 
(An important exception to this is when $n_0$ is a half-integer. )
It is possible to actually observe changes in the behaviour of the 
conductance as $n_0$ is varied by a single step.  For $n_0$ close 
to an integer value the conductance tends to become small at low $T$. 
This is a consequence of the fact that, for an electron to pass 
through the dot it must go temporarily into a high energy state with 
$n=n_0\pm 1$, an effect known as the Coulomb blockade.  At the special 
values of the gate voltage where $n_0$ is a half-integer, the Coulomb 
blockade is lifted and the conductance is larger.  
\subsection{Single channel Kondo effect}
At still lower temperatures, a difference emerges between the case where $n_0$
is close to an even or odd integer; this is due to the Kondo effect. When 
$n_0$ is near an odd integer the dot must have a non-zero (half-integer) spin, 
generally $1/2$. At energy scales small compared to $U$, we may disregard 
charge fluctuations on the dot and consider only its spin degrees of freedom. 
Virtual processes, of second order in $t$, lead to a Kondo exchange interaction 
between the spin on the quantum dot and the spin of the mobile electrons 
on the left and right side of the 2DEG.  A simplified and well-studied 
model is obtained by considering only a single energy level on the quantum 
dot, the one nearest the Fermi energy. Then there are only four states 
available to the quantum dot: zero or two electrons or one electron 
with spin up or down. The corresponding model is known as the Anderson Model (AM):
\be H = \int dk \psi^{\dagger\alpha}_k\psi_{k\alpha}\epsilon (k) +\Gamma 
\int dk 
[\psi^{\dagger\alpha}_kd_\alpha +h.c.]+{U\over 2}(\hat n_d-n_0)^2,\ee
where 
\be \hat n_d\equiv d^{\dagger \alpha}d_\alpha .\ee
$d^{\dagger\alpha}$ creates an electron in the single energy level under 
consideration on the dot. Note that I am now treating the conduction 
electrons as one-dimensional. This is motivated by the fact 
that the point contacts between the 2DEG regions and the dot 
are assumed to be single-channel. However, the actual 
wave-functions of the electrons created by $\psi^{\dagger \alpha}_k$
 are extended in two dimensions on the left and right sides of the dot.
The conventional label $k$, doesn't really label a wave-vector anymore. 
An important assumption is being made here that there is a set 
of energy levels, near the Fermi energy, which are equally spaced 
and have equal hybridization amplitudes, $t$, with the $d$-level 
on the quantum dot. This is expected to be reasonable for 
small quantum dots with weak tunnelling amplitudes, $t$ and smooth 
point contacts. I assume, for convenience, that these wave-functions are parity symmetric 
between the left and right 2DEG's. As the gate voltage 
is varied, $n_0$ passes through $1$.  Provided that 
$\Gamma^2\nu \ll U$, where $\nu$ is the (1-dimensonal) density of states, we 
can obtain the Kondo model as the low energy effective theory 
at scales small compared to $U$, with an effective Kondo coupling:
\be J={2\Gamma^2\over U(2n_0-1)(3-2n_0)}.\ee
Thus we again expect the spin of the quantum dot to be 
screened at $T\ll T_K$ by the conduction electrons in the leads. 

A crucial, and perhaps surprising point is how the Kondo physics 
affects the conductance through the dot. This is perhaps 
best appreciated by considering a tight binding version 
of the model, where we replace the left and right leads 
by 1D tight-binding chains:
\be H=-t\sum_{j=-\infty}^{-2}(c^\dagger_jc_{j+1}+h.c.]
-t\sum_{1}^{\infty}(c^\dagger_jc_{j+1}+h.c.)
-t'[(c^\dagger_{-1}+c^\dagger_{1})d+h.c.)]+{U\over 2}(\hat n_d-n_0)^2.\ee
The Kondo limit gives:
\be  H=-t\sum_{j=-\infty}^{-2}(c^\dagger_jc_{j+1}+h.c.)
-t\sum_{1}^{\infty}(c^\dagger_jc_{j+1}+h.c.)
+J(c^\dagger_{-1}+c^\dagger_1){\vec \sigma \over 2}(c_{-1}+c_1)\cdot \vec S\ee
with
\be J={2t'^2\over U(2n_0-1)(3-2n_0)}.\ee
Note that the only way electrons can pass from the right to the 
left lead is via the Kondo interaction. Thus if the Kondo 
interaction is weak, the conductance should be small. The 
renormalization of the Kondo coupling to large values at 
low energy scales implies a dramatic characteristic {\it increase}
of the conductance upon lowering the temperature. In the 
particle-hole symmetric case of half-filling, it is easy 
to understand the low temperature limit by simply taking 
the bare Kondo coupling, to infinity $J\gg t$. Now the 
spin of the quantum dot (at site $0$) forms a singlet 
with an electron in the parity-symmetric state on sites $1$ and $-1$:
\be \left[ d^{\dagger\uparrow}
{(c_{-1}^{\dagger \downarrow}+c_1^{\downarrow})\over \sqrt{2}}
 -d^{\dagger\downarrow}
{(c_{-1}^{\dagger \uparrow}+c_1^{\uparrow})\over \sqrt{2}}\right]|0>.\ee
The parity-antisymmetric orbital
\be c_a\equiv {c_{-1}-c_1\over \sqrt{2}},\ee
is available to conduct current past the screened dot. 
The resulting low-energy effective Hamiltonian:
\be  H=-t\sum_{j=-\infty}^{-3}(c^\dagger_jc_{j+1}+h.c.)
-t\sum_{2}^{\infty}(c^\dagger_jc_{j+1}+h.c.)
-{t\over \sqrt{2}}[(-c^\dagger_{-2}+c_2)c_a+h.c.],\ee
has resonant transmission, $T_r=1$ at the Fermi energy in the particle-hole symmetric
 case of half-filling. This leads to ideal $2e^2/h$ conductance from the 
Laudauer formula. Thus we expect the conductance to 
increase from a small value of order $J^2$ at $T\gg T_K$ to the 
ideal value at $T\ll T_K$. Thus, the situation is rather inverse to 
the case of the resistivity due to a dilute random array of Kondo scatterers 
in 3 (or lower) dimensions.  For the quantum dot geometry discussed here 
lowering $T$ leads to an increase in conductance rather than 
an increase in resistivity. Actually, it is easy to find another 
quantum dot model, referred to as ``side-coupled'' where the 
behaviour is like the random array case. In the side coupled 
geometry the tight-binding Hamiltonian is:
\be  H=-t\sum_{j=-\infty}^{\infty}(c^\dagger_jc_{j+1}+h.c.)
+Jc^\dagger_0{\vec \sigma \over 2}c_0\cdot \vec S.\ee
Now there is perfect conductance at $J=0$ due to the direct 
hopping from sites $-1$ to $0$ to $1$. On the other hand, 
in the strong Kondo coupling limit, an electron sits at site $0$ 
to form a singlet with the impurity. This completely blocks 
tranmission since an electron cannot pass through without 
destroying the Kondo singlet. 

At arbitrary temperatures, the conductance throught the quantum dot may be 
expressed exactly in terms of the ${\cal T}$-matrix, ${\cal T}(\omega ,T)$.  This is 
precisely the same function which determines the resistivity for a dilute random
 array of Kondo scatterers. To apply the Kubo formula, it is important 
to carefully distinguish conduction electron states in the left 
and right leads. Thus we write the Kondo Hamiltonian in the form:
\be H= \sum_{L/R}\int dk \psi^{\dagger}_{L/R,k}\psi_{L/R,k}\epsilon (k)
+{J\over 2}\int {dk dk'\over 2\pi}(\psi^\dagger_{L,k}+\psi^\dagger_{R,k}){\vec \sigma \over 2}
(\psi_{L,k'}+\psi_{R,k'})\cdot \vec S.
\label{HKQD}\ee
The Kondo interaction only involves the symmetric combination of left and right leads. 
On the other hand, the current operator, which appears in the Kubo formula for 
the conductance, is:
\be j=-e{d\over dt}[N_L-N_R]=-ie[H,N_L-N_r],\ee
where $N_{L/R}$ are the number operators for electrons in the left and right leads:
\be N_{L/R}\equiv \int dk \psi^\dagger_{L/R,k}\psi_{L/R,k}.\ee
Introducing symmetric and anti-symmetric combinations:
\be \psi_{s/a}\equiv {\psi_L\pm \psi_R\over \sqrt{2}},\ee
the Kondo interaction only involves $\psi_s$ but the current operator is:
\be j={d\over dt}\int dk[\psi^\dagger_s\psi_a+h.c.],\ee
which contains a product of symmetric and anti-symmetric operators. 
The Kubo formula: 
\be G=\lim_{\omega \to 0}{1\over \omega}\int_0^\infty e^{i\omega t}<[j(t),j(0)]>,\ee
then then be expressed as a product of the free Green's function for $\psi_a$ 
and the interacting one for $\psi_s$. Expressing the $\psi_s$ Green's 
function in terms of the ${\cal T}$-matrix, by Eq. (\ref{Tmatrix}), it is 
not hard to show that the conductance is given by:
\be G(T)={2e^2\over h}\int d\epsilon \left[-{dn_F\over d\epsilon}(T)\right]
[-\pi \nu \hbox{Im}{\cal T}(\epsilon ,T)].\ee
For $T\gg T_K$ a perturbative calculation of the ${\cal T}$-matrix 
gives:
\be -2\pi \nu {\cal T}\to -{3\pi^2\over 8}\lambda^2+\ldots \ee
It can be checked that the higher order terms replace the Kondo coupling, 
$\lambda$, by its renormalized value at scale $\omega$ of $T$ (whichever is higher), 
leading to the conductance:
\be G\to {2e^2\over h}{3\pi^2\over 16 \ln^2(T/T_K)},\ \  (T\gg T_K).\ee
On the other hand, at $T$, $\omega \to 0$, $-2\pi \nu {\cal T}\to -2i$, 
corresponding to the $\pi /2$ phase shift, leading to ideal conductance. 
By doing second order perturbation theory in the LIO, Nozi\`eres Fermi liquid theory 
gives:
\be G\to {2e^2\over h}\left[ 1-\left({\pi^2 T\over 4T_K}\right)^2\right] .\ee

The calculation of ${\cal T}$ at intermediate temperatures and 
frequencies of order $T_K$ is a difficult problem. It goes 
beyond the scope of our RG methods which only apply near 
the high and low energy fixed points. It is also 
not feasible using the Bethe ansatz solution of the Kondo model. 
The most accurate results at present come from the 
Numerical Renormalization Group method. 

\subsection{Two channel Kondo effect}
Quite recently, the first generally accepted experimental realization 
of an overscreened Kondo effect, in the two-channel, $S=1/2$ case, 
was obtained in a quantum dot device. To realize the difficulty 
in obtaining a two-channel situation, consider the case discussed 
above. In a sense there are two channels in play, corresponding 
to the left and right leads. However, the problem is that 
only the even channel actually couples to the spin of the quantum dot. 
In Eq. (\ref{HKQD}) it is the left-right cross terms in the Kondo interaction 
that destroy the two-channel behaviour.  If such terms could somehow 
be eliminated, we would obtain a two-channel model. On the other 
hand, the only thing which is readily measured in a quantum dot 
experiment is the conductance, and this is trivially zero 
if the left-right Kondo couplings vanish. 

\begin{figure}[th]
\begin{center}
\includegraphics[clip,width=8cm]{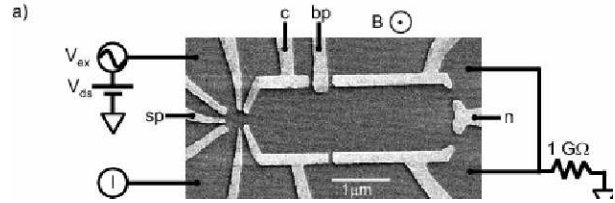}
\caption{Device for realizing the 2-channel Kondo effect.}
\label{fig:GG}
\end{center}
\end{figure}

A solution to this problem, proposed by Oreg and Goldhaber-Gordon\cite{Oreg}, 
involves a combination of a small dot, in the Kondo regime, 
and a ``large dot'', i.e. another, larger puddle of conduction 
electrons with only weak tunnelling from it to the rest of the system, as shown 
in figure (\ref{fig:GG}). This was then realized experimentally 
by Goldhaber-Gordon's group.\cite{Potok} The key feature is to adjust the size of 
this large dot to be not too large and not too small. It should 
be chosen to be large enough that the finite size level spacing 
is negligibly small compared to the other relevant energy scales:
$T_K$ and $T$ which may be in the mili-Kelvin to kelvin range. 
On the other hand, the charging energy of the dot, effectively
 the $U$ parameter discussed above, must be relatively 
large compared to these other scales. In this case, 
the charge degrees of freedom of the large dot are frozen 
out at low energy scales. An appropriate Kondo type model could be written as:
\be H= \sum_{j=1}^3\int dk \psi^{\dagger}_{j,k}\psi_{j,k}\epsilon (k)
+{2\over U_s}\sum_{i,j=1}^3\Gamma_i\Gamma_j
\int {dk dk'\over 2\pi}\psi^\dagger_{i,k}{\vec \sigma \over 2}
\psi_{j,k'}\cdot \vec S+{U_l\over 2}(\hat n_3-n_0)^2.
\label{QD2C}\ee
Here $j=1$ corresponds to the left lead, $j=2$ corresponds to the right lead 
and $j=3$ corresponds to the large dot. $U_{s/l}$ is the charging energy 
for the small/large dot respectivly, $\Gamma_i$ are 
the corresponding tunelling amplitudes onto the small dot
 $\hat n_3$ being the total number of electrons 
on the large dot. $n_0$ is the lowest energy electron number for the large 
dot, which is now a rather large number. (It is actually unimportant 
here whether $n_0$ is an integer or half-integer because the 
Kondo temperature for the large dot is assumed to be negligibly small.) 
If $U_l$ is sufficiently large, the 
$1$-$3$ and $2$-$3$ cross terms in the Kondo interaction 
can be ignored, since they takes the large dot from a low energy state 
with $n_3=n_0$ to a high energy state with $n_3=n_0\pm 1$. Dropping 
these cross terms, assuming $\Gamma_1=\Gamma_2$ as before, 
and replacing $(\psi_1+\psi_2)/\sqrt{2}$ by $\psi_s$, as 
before, we obtain a two-channel Kondo model, but with 
different Kondo couplings for the two channels:
\bea J_1&\equiv& {4\Gamma_1^2\over U_s}\nonumber \\
J_2&\equiv& {2\Gamma_3^2\over U_s}.\eea
Finally, by fine-tuning $\Gamma_3$ it is possible 
to make the two Kondo couplings equal, obtaining 
precisely the standard two-channel Kondo Hamiltonian. 

Using our BCFT methods it is easily seen that this type 
of channel anisotropy, with $J_1\neq J_2$, is a relevant perturbation.\cite{Affleck5}
The relevant operator which now appears at the low energy fixed point is 
\be \delta H \propto (J_1-J_2)\varphi^3_c.\ee
Here $\vec \varphi_c$ is a primary field in the channel sector.  
Since the associated channel WZW model is also $SU(2)_2$ for $k=2$ 
we may label channel fields by their corresponding pseudo-spin. This 
primary field has pseudo-spin 1, and scaling dimension $1/2$. 
It must be checked that it occurs in the boundary operator spectrum 
at the Kondo fixed point. This follows from the fact 
that $\varphi_c^a\varphi_s^b$, the product of channel and spin 
$j=1$ primaries occurs for free fermion BC's. This dimension 1 
operator occurs in the non-abelian bosonization formula:
\be \psi_L^{\dagger j\alpha}(\sigma^a)_\alpha^\beta (\sigma^b)_j^i\psi_{L i\beta}
\propto \varphi^a_s\varphi^b_c.\ee
(Note that both sides of this equation have the same scaling dimensions 
and the same symmetries.) We get the boundary operator spectrum 
at the Kondo fixed point by double fusion with the $j=1/2$ primary 
in the spin sector. The first fusion operation maps 
the $j_s=1$ spin primary into $j_s=1/2$ and the second one 
maps $j_s=1/2$ into $j_s=0$ (and $j_s=1$). Therefore, the 
$j_c=1$, $j_s=0$ primary is in the boundary operator 
spectrum at the (overscreened) Kondo fixed point. Since 
this operator exists and is allowed by all symmetries 
once the channel $SU(2)$ symmetry is broken, we expect 
it to be generated in the low energy effective Hamiltonian. 
It thus destablizes the fixed point since it is relevant. 
It is not hard to guess what stable fixed point the system 
flows to.  Suppose $J_1>J_2$. The stable fixed point corresponds 
to $J_2$ flowing to zero and $J_1$ flowing to large values. 
The more strongly coupled channel $1$ screens the $S=1/2$ impurity 
while the more weakly coupled channel $2$ decouples. This 
fixed point is characterized by simple phase shifts of $\pi /2$ 
for channel $1$ and $0$ for channel $2$. Such behaviour 
is consistent with the weak coupling RG equations:
\bea {d\lambda_1\over d \ln D}&=&-\lambda_1^2+{1\over 2}\lambda_1(\lambda_1^2+\lambda_2^2)
+\ldots \nonumber \\
{d\lambda_2\over d \ln D}&=&-\lambda_2^2+{1\over 2}\lambda_2(\lambda_1^2+\lambda_2^2)
+\ldots .\eea
Once $\lambda_1^2$ gets larger than $2\lambda_2$, these equations 
predict that the growth of $\lambda_2$ is arrested and it starts to decrease. 
This RG flow is also consistent with the $g$-theorem: $g=(1/2) \ln 2$ at 
the  symmetric fixed point but $g=0$ at the stable fixed point. 
The implications of this RG flow for the conductance 
in the quantum dot system is also readily deduced.\cite{Pustilnik} From Eq. (\ref{T0}) 
using $S_{(1)}=0$ for the $k=2$, $S=1/2$ Kondo fixed point, we see 
that ${\cal T}(\omega=T=0)$ has half the value it has in the Fermi liquid case, 
and therefore the conductance through the quantum dot has half 
the value for the single-channel fixed point, $G(0)=e^2/h$. 
Let's call the Kondo coupling to the large dot $J_l$ and 
the coupling to the symmetric combination of left and right leads, $J_s$. The $T=0$ 
conductance when $J_s>J_l$ is $2e^2/h$, due to the $\pi /2$ phase shift 
in the ${\cal T}$-matrix for the $s$ channel. On the other hand, 
if $J_l>J_s$, the $T=0$ conductance is zero since the $s$-channel 
phase shift is zero. For bare couplings that are close to each other, 
$\lambda_1\approx \lambda_2$ the system will flow towards the 
NFL critial point, before diverging from it at low $T$. i.e. there is a 
``quantum critical region'' at finite $T$ for $\lambda_1$ near $\lambda_2$. 
The basic scaling properties follow from the fact the relevant operators 
destabilizing the NFL critical point has dimension 1/2, together 
with the fact that the LIO at the NFL critical point has dimension 3/2.
Right at the critical point, the finite $T$ correction to the conductance, 
of first order in the LIO is:
\be G(T)\to {e^2\over h}[1-(\pi T/T_K)^{1/2}].\ee
(The prefactor  was also determined exactly here, but this is only useful 
if some independent measure of $T_K$ can be made experimentally.) We 
define $T_c$ as the crossover scale, at which the RG flow starts to 
deviate from the NFL critical point.  This defines the energy 
scale occuring in the relevant perturbation in the effective Hamiltonian:
\be H=H_{NFL}\pm T_c^{1/2}\varphi^3_c(0)-{1\over T_K^{1/2}}\vec J_{-1}\cdot \vec \varphi_s,\ee
where I have included both the LIO (last term) and the relevant operator, 
which is present when $\lambda_1\neq \lambda_2$ with a sign for the coupling constant 
$\propto \lambda_1-\lambda_2$. 
For almost equal bare couplings, 
this will be $\ll T_K$, the scale at which the renormalized couplings become large. 
At $T_c\ll T\ll T_K$, we can calculate the correction to the NFL conductance 
to first order in $\varphi^c_c$, giving:
\be G(T)\approx {e^2\over h}[1+\hbox{constant}\cdot \hbox{sgn} (\Delta )(T_c/T)^{1/2}].\ee
Here:
\be \Delta \equiv \lambda_1-\lambda_2.\ee

We may also estimate\cite{Pustilnik} $T_c$ in terms of $\Delta$ and the average bare Kondo coupling:
\be \bar \lambda \equiv {\lambda_1+\lambda_2\over 2}.\ee
The weak coupling RG  equations (to second order only) are:
\bea {d \bar \lambda\over d\ln D}&=&-\bar \lambda^2\nonumber \\
{d\Delta \over d\ln D}&=&-2\Delta \bar \lambda .\eea
The solution to the first of these can be written:
\be \bar \lambda (D)={1\over \ln (D/T_K)},\ee
for $D\gg T_K$. The second of these RG equations then can be written:
\be {d \Delta \over d\ln D}=-{2\over \ln (D/T_K)}\Delta \ee
Integrating this equation gives:
\be \Delta (T_K)\propto \Delta_0/\bar \lambda_0^2,\label{DTK}\ee
where $\bar \lambda_0$ and $\Delta_0$ are the bare couplings. 
At energy scales below $T_K$ we may write the RG equation for $\Delta (D)$:
\be {d\Delta \over d ln D}={1\over 2}\Delta ,\ee
reflecting the fact that $\Delta$ has scaling dimension $1/2$ at the NFL fixed point. 
Thus:
\be \Delta (D)=\left({T_K\over D}\right)^{1/2}\Delta (T_K)\ee
for $D<T_K$. 
By definition, the crossover scale, $T_c$ is the energy scale where $\Delta (D)$ 
becomes of order 1.  Thus:
\be 1\propto \Delta (T_K)\left({T_K\over T_C}\right)^{1/2}.\ee
Using Eq. (\ref{DTK}) for $\Delta (T_K)$, we finally determine the crossover 
scale in terms of $T_K$ and the bare parameters:
\be T_c\propto T_K{\Delta_0^2\over \bar \lambda_0^4}.\ee

Another interesting quantity is the dependence of the conductance, at $\Delta_0=0$ 
on the temperature and the source-drain voltage. $V_{sd}$ defines another 
energy scale, in addition to $T$ so we expect:
\be G\equiv {dI\over dV_{sd}}={e^2\over h}\left[ 1-\left({\pi T\over T_K}\right)^{1/2}F(eV_{sd}/T)\right].\ee
where $F(x)$ is some universal scaling function.  A theoretical calculation of $F$ 
remains an open problem.  These theoretical predictions, in particular the 
occurance of the critical exponent 1/2, are in good agreement with 
the experiments of the Goldhaber-Gordon group.\cite{Potok}

\section{Quantum Impurity problems in Luttinger liquids}
The Kondo models considered so far in these lectures all have the property that the 
electrons are assumed to be non-interacting, except with the impurity. The validity 
of this approximation, is based on Fermi liquid theory ideas, as mentioned in the first lecture. 
Although our model become 1 dimensional after s-wave projection, it is probably 
important that it was originally 2 or 3 dimensional, to justify ignoring these interactions, 
since in 1 dimensional case, Fermi liquid theory definitely fails. Now interactions 
are important leading, at low energies, to ``Luttinger liquid'' (LL)  behaviour.  We 
will now find interesting boundary RG phenomena for a potential scatterer, 
even without any dynamical degrees of freedom at the impurity.\cite{Kane,Eggert} The physical 
applications of this theory include a point contact in a quantum wire, or a carbon nano-tube, a constriction 
in a quantum Hall bar or impurities in spin chains. 

I will just give a lightening review of LL theory here, since it has been reviewed 
many other places (for example [\onlinecite{Affleck0,Giamarchi}]) and is not the main subject of these lectures. We are generally 
interested in the case of fermions with spin, but no additional ``channel'' quantum 
numbers ($k=1$). A typical microscopic model is the Hubbard model:
\be H=-t\sum_j[(\psi^\dagger_j\psi_{j+1}+h.c.)+U\hat n_j^2]\ee
where $\hat n_j$ is the total number operator (summed over spin directions) on site $j$. 
(More generally, we might consider ``ladder'' models in which case we would also 
get several ``channels'' and a plethora of complicated interactions.) 
Non-abelian bosonization is again useful, leading to a separation 
of spin and charge degrees of freedom. But now we must consider the various 
bulk interactions. These fall into several classes:
\begin{itemize}
\item{$g_cJ_LJ_R$: An interaction term of this form,  
which is proportional to $(\partial_\mu \varphi )^2$,  in the Lagrangian density, can be eliminated 
by rescaling the charge boson field: $\varphi \to \sqrt{g}\varphi$. Here the Luttinger parameter, $g$, 
has the value 
$g=1$ in the non-interacting case.  (Unfortunately there are numerous different 
conventions for the Luttinger parameter.  I follows here the notation of [\onlinecite{Oshikawa}].) This leaves 
the Hamiltonian in non-interacting form, but the rescaling 
changes the scaling dimensions of various operators.}
\item{$\psi^{\dagger \uparrow}_R\psi^{\dagger \downarrow}_R\psi_{L\uparrow}\psi_{L\downarrow}$+h.c.:
This can be bosonized as a pure charge operator.  Depending on the value of the Luttinger parameter,  
it can be relevant in which case it produces a gap for charge excitations.  However, 
this ``Umklapp'' term is accompanied by oscillating factors $e^{\pm 2ik_Fx}$ so it can 
usually be ignored unless $k_F=\pi /2$, corresponding to half-filling. It 
produces a charge gap in the repulsive Hubbard model at half-filling. The low energy 
Hamiltonian then involves the spin degrees of freedom only.  In particular, it may 
correspond to the $SU(2)_1$ WZW model.  In the large $U$ limit of the Hubbard model, 
we obtain the S=1/2 Heisenberg model, with antiferromagnetic coupling $J\propto t^2/U$, 
as a low energy $E\ll U$) lattice model. The low energy Hamiltonian for the Heisenberg 
model is again the $SU(2)_1$ WZW model.}
\item{Marginal terms of non-zero conformal spin.  The only important affect of these 
is assumed to be to change the velocities of spin and charge degrees of freedom, 
making them, in general, different.}
\item{$-(g_s/2\pi )\vec J_L\cdot \vec J_R$: $g_s$ has a quadratic $\beta$-function 
at weak coupling; it flows to zero logarithimically if it is initially positive, as occurs 
for the repulsive, $U>0$ Hubbard model. 
It is often simply ignored, but, in fact, it leads to important logarithmic 
corrections to all quantities.}
\item{Spin anisotropic interactions of zero conformal spin: Often $SU(2)$ spin 
symmetry is a good approximation in materials but it is generally broken to some extent, 
due to spin-orbit couplings. If a $U(1)$ spin symmetry is preserved then, 
depending on parameters, the spin degrees of freedom can remain gapless. 
It is then usually convenient to use ordinary abelian bosonization. The spin 
boson then also also gets rescaled $\varphi_s\to g_s\varphi_s$ where 
$g_s=1$ in the isotropic case. This leads to further changes in scaling dimensions 
of various operators.}
\item{Various higher dimensional operators of non-zero conformal spin: Some 
of these have very interesting and non-trivial effects and are  the 
subject of current research. However, these effects generally go away at low 
energies.}
\end{itemize}
Let us begin with an interacting spinless fermion model with impurity scattering 
at the origin only, corresponding to a point contact in a quantum wire. A corresponding lattice model would be, for example:
\be H=[-t\sum_{j=-\infty}^{-1}\psi^\dagger_j\psi_{j+1}-t'\psi^\dagger_0\psi_{1}
-t\sum_{j=1}^{\infty}\psi^\dagger_j\psi_{j+1}+h.c.]
+U\sum_{j=-\infty}^\infty \hat n_j\hat n_{j+1}.\label{fxxz}\ee
The hopping term between sites $0$ and $1$ has been modified from $t$ to $t'$; we might expect $t'\ll t$ 
for a point contact or constriction. 
Upon bosonizing and rescaling the boson, the bulk terms in the action just give:
\be S_0={g\over 4\pi}\int_{-\infty}^\infty dxd\tau (\partial_\mu \varphi )^2 .\ee
The impurity term, in terms of continuum limit fermions, 
\be \sqrt{2\pi} \psi_j\approx  e^{ik_Fj}\psi_R(j)+e^{-ik_Fj}\psi_R(j),\label{fermCL}\ee
is:
\be H_{int}\approx {t-t'\over 2\pi}[J_L(0)+J_R(0)+(\psi^\dagger_L(0)\psi_R(0)e^{ik_F}+h.c.)].\ee
(Note that we ignore the small variation of the continuum limit fields over one lattice spacing here. 
Including this effect only leads to irrelevant operators. This continuum limit Hamiltonian 
is appropriate for small $|t'-t|$, since we have taken the continuum limit 
assuming $t'=t$.)
Using the bosonization formulas:
\be \psi_{L/R}\propto e^{i(\varphi \pm \theta )/\sqrt{2}}\nonumber \\
\label{fermbos}\ee
 this becomes:
\be H_{int} =-(t'-t)\sqrt{2}\partial_x\theta (0) -\hbox{constant}\cdot(t'-t)\cos [\sqrt{2}(\theta (0)-\alpha )],\ee
for a constant $\alpha$ depending on $k_F$.
 While the first term is always exactly marginal, 
the second term, which arises from ``backscattering'' ($L\leftrightarrow R$) has 
dimension
\be x=g.\ee
It is marginal for free fermions, where $g=1$ but is relevant for $g<1$, 
corresponding to repulsive interactions, $U>0$. It is 
convenient to go a basis of even and odd channels. 
\be \theta_{e/o}(x)\equiv {\theta (x)\pm \theta (-x)\over \sqrt{2}}\ee
 
The $\theta_{e/o}$ fields obey Neumann (N) and Dirichlet (D) BC's respectively:
\bea \partial_x\theta_e(0)&=&0\nonumber \\
\theta_o(0)&=&0.\eea
Then the action separates into even and odd parts, $S=S_e+S_o$ with:
\bea S_e&=&{1\over 4\pi g}\int_{-\infty}^\infty d\tau \int_0^\infty dx(\partial_\mu \theta_e )^2-V_b\cos 
(\theta_e(0)-\alpha )\nonumber \\
S_o&=&{1\over 4\pi g}\int_{-\infty}^\infty d\tau \int_0^\infty dx (\partial_\mu \theta_o )^2-V_f\partial_x\theta_o(0),\eea
where $V_{f/b}$, the forward and backward scattering amplitudes, are both $\propto t'-t$.  The interaction 
term can be eliminated from $S_o$ by the transformation:
\be \theta_o(x)\to \theta_o(x)-2\pi V_fg\cdot \hbox{sgn}(x).\ee
On the other hand, $S_e$ is the well-known boundary sine-Gordon model which is not so easily solved. 
It is actually integrable\cite{Ghoshal} and a great deal is known about it, but here I will just discuss simple RG results. 
For $g<1$, when backscattering is relevant, it is natural to assume that $V_b$ renormalizes to infinity 
thus changing the N boundary condition on $\theta_e$ to D, $\theta_e(0)=\alpha$. This has the effect 
of severing all communication between left and right sides of the system, corresponding to a cut chain. 
If the forward scattering, $V_f=0$ then we have independent D boundary conditions on left and right sides:
\be \theta (0^\pm )=\alpha /\sqrt{2}.\label{Dsimp}\ee
For non-zero $V_f$, left and right side are still severed but the D BC's are modified to:
\be \theta (0^{\pm})=\alpha /\sqrt{2}\mp \sqrt{2}\pi V_fg.\label{DV}\ee
The simple D BC of Eq. (\ref{Dsimp}) or (\ref{DV}) correspond, in the original fermion language to:
\be \psi_L(0^\pm )\propto \psi_R(0^\pm ).\ee
The right moving excitations on the $x<0$ axis are reflected at the origin picking up a phase shift 
which depends on $V_f$, and likewise for the left moving excitations on $x>0$.

Of course, we have made a big assumption here that $V_b$ renormalizes to $\infty$ giving 
us this simple D BC.  It is important to at least check the self-consistency of the 
assumption. This can be done by checking the stability of the D fixed point. 
Thus we consider the Hamiltonian of Eq. (\ref{fxxz}) 
with $t'\ll t$. To take the continuum limit, we must carefully take into account the boundary 
conditions when $t'=0$. Consider the chain from $j=1$ to $\infty$ with open boundary conditions (OBC). 
This model is equivalent to one where a hopping term, of strength $t$, to site $0$ is included 
but then a BC $\psi_0=0$ is imposed. From Eq. (\ref{fermCL}) we see that this corresponds to 
$\psi_L(0)=-\psi_R(0)$.  Using the bosonization formulas of Eq. (\ref{fermbos}) we see that 
this corresponds to a D BC, $\theta (0)=$ constant, as we would expect from the previous 
discussion. A crucial point is that imposing a D BC changes the scaling dimension 
of the fermion fields at the origin.  Setting $\theta (0)=$ constant, Eq. (\ref{fermbos}) reduces to:
\be \psi_{L/R}(0)\propto e^{i\varphi (0)/\sqrt{2}}.\label{fermD}\ee
The dimension of this operator is itself affected by the D BC. Decomposing $\varphi (t,x)$ and $\theta$ into 
left and right moving parts:
\bea \varphi (t,x)&=&{1\over \sqrt{g}}[\varphi_L(t+x)+\varphi_R(t-x)]\nonumber \\
\theta &=&\sqrt{g}[\varphi_L-\varphi_R]
\eea
we see that the D BC implies:
\be \varphi_R(0)=\varphi_L(0)+\hbox{constant}.\ee
Thus to evaluate correlation functions of $\varphi (0)$ with D BC we can use:
\be \varphi (0)\to {2\over \sqrt{g}}\varphi_L(0)+\hbox{constant}.\ee
The bulk correlation functions of exponentials of $\varphi$ decay as:
\be <e^{ia\varphi (t,x)}e^{-ia\varphi (0,0)}>=<e^{ia\varphi_L(t+x)/\sqrt{g}}e^{-ia\varphi_L(0,0)/\sqrt{g}}>\cdot 
<e^{ia\varphi_R(t-x)/\sqrt{g}}e^{-ia\varphi_R(0,0)/\sqrt{g}}>
={1\over (x+t)^{a^2/2g}(x-t)^{a^2/2g}}
\ee
On the other hand, at a boundary with a D BC,
\be <e^{ia\varphi (t,0)}e^{-ia\varphi (0,0)}>=<e^{2ia\varphi_L(t)/\sqrt{g}}e^{-2ia\varphi_L(0,0)/\sqrt{g}}>
={1\over t^{(2a)^2/2g}}.\ee
The RG scaling dimension of the operator $e^{ia\varphi }$ doubles at a boundary with a D BC to 
$\Delta = a^2/g$. Thus the fermion field at a boundary with D BC, Eq. (\ref{fermD}), has 
a scaling dimension $1/2g$. The weak tunnelling amplitude $t'$ in Eq. (\ref{fxxz}) couples 
together two independent fermion fields from left and right sides, both obeying D BC's.  Therefore 
the scaling dimension of this operator is obtained by adding the dimension of each 
independent fermion fields, and has the value $1/g$. This is relevant when $g>1$, 
the case where the weak backscattering is irrelevant, and is irrelevant for $g<1$ the 
case where the weak backscattering is relevant. Thus our bold conjecture that 
the backscattering, $V_b$ renormalizes to $\infty$ for $g<1$ has passed an important consistency test. 
The infinite back-scattering, cut chain, D BC fixed point is indeed stable for $g<1$. On the 
other hand, and perhaps even more remarkably, it seems reasonable to hypothesize 
that even a weak tunnelling $t'$ between two semi-infinite chains flows 
to the N BC at low energies.  This is a type of ``healing'' phenomena: Translational 
invariance is restored in the low energy, long distance limit. 

The conductance is clearly zero at the D fixed point.  At the N fixed point we may 
calculate it using a Kubo formula. One approach is to apply an AC electric field 
to a finite region, $L$, in the vicinity of the point contact:
\be G=\lim_{\omega \to 0}{-e^2\over h}{1\over \pi \omega L}\int_{-\infty}^\infty d\tau e^{i\omega \tau}
\int_0^LdxT<J(y,\tau )J(x,0)>,\ee
(independent of $x$). 
Here the current operator is $J=-i\partial_\tau \theta $. Using:
\be <\theta (x,\tau )\theta (0,0)>=-{g\over 2}\ln (\tau^2+x^2),\ee
for the infinite length system, 
it is straightforward to obtain
\be G=g{e^2\over h}.\ee
(While this is the conductance predicted by the Kubo formula, it is apparently 
not neccessarily what is measured experimentally.\cite{Tarucha}  For various theoretical discussions of this point see [\onlinecite{gG}].)
Low temperature corrections to this conductance can be obtained by doing perturbation theory in the LIO, 
as usual. At the cut chain, D, fixed point, for $g<1$, the LIO is the tunnelling term, $\propto t'$ in Eq. (\ref{fxxz}). 
This renormalizes as:
\be t'(T)\approx {t'}_0(T/T_0)^{1/g-1}.\ee
Since the conductance is second order in $t'$ we predict:
\be G(T)\propto {t'}_0^2T^{2(1/g-1)},\ \  (g<1, T\ll T_0).\ee
Here $T_0$ is the lowest characteristic energy scale in the problem.  If the bare ${t'}_0$ is small 
then $T_0$ will be of order the band width, $t$.  However, if the bare model has only a weak 
back-scattering $V_b$ then $T_0$ can be much smaller, corresponding to the energy scale 
where the system crosses over between N and D fixed points, analgous to the Kondo temperature. 
(Note that there is no Kondo impurity spin in this model, however.) 
On the other hand, near the N fixed point, where the backscattering is weak we may 
do perturbation theory in the renormalized back scattering; again the contribution to $G$ is 
second order. Now, for $g>1$, 
\be V_b(T)=V_{b0}(T/T_0)^{g-1}\ee
and hence:
\be G-ge^2/h\propto V_b^2T^{2(g-1)},\ \  (g>1, T\ll T_0).\ee
Again $T_0$ is the lowest characteristic energy scale; it is a small cross over scale 
if the microscopic model has only a small tunnelling $t'$. 

\begin{figure}[th]
\begin{center}
\includegraphics[clip,width=8cm]{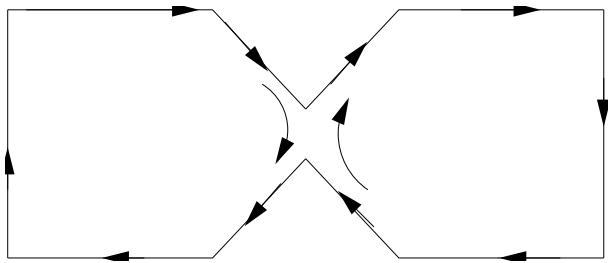}
\caption{Quantum Hall bar with a constriction. Edge currents circulate 
clockwise and can tunnel from upper to lower edge at constriction.}
\label{fig:Hall}
\end{center}
\end{figure}

A beautiful application\cite{Moon,Fendley} of this quantum impurity model is to tunnelling through a 
constriction in a quantum Hall bar, as illustrated in fig. (\ref{fig:Hall}). 
Consider a 2DEG in a strong magnetic field at the fractional quantum Hall effect plateau 
of filling factor $\nu =1/3$. Due to the bulk excitation gap in the Laughlin ground state 
there is no current flowing in the bulk of the sample.  However, there are gapless edge 
states which behave as a chiral Luttinger liquid. 
Now the currents are chiral with right movers restricted to the lower edge 
and left movers to the upper edge in the figure. Nonetheless, we may apply our 
field theory to this system and the Luttinger parameter turns out 
to have the value $g=\nu <1$.  (Furthermore, the edge states are 
believed to be spin-polarized, making the spinless model discussed here appropriate. )
Right and left movers interact at the 
constriction, with a finite probability of back scattering, which in this 
case takes quasi-particles between upper and lower edges. The entire 
cross over function for the conductance can be calculated, either by quantum 
Monte Carlo (together with a delicate analytic continuation to zero frequency) 
or using the integrability of the model.  The results agree fairly well with experiments. 

A number of other interesting quantum impurity problems have been studied in Luttinger liquids. 
These include the generalization of the model discussed above to include electron spin.\cite{Kane} 
Four simple fixed points are found which are obvious generalizations of the two 
discussed in the spinless case.  We may now have D or N BC for both charge and spin 
bosons, corresponding to perfect reflection/transmission for charge/spin. Interestingly, 
additional fixed points occur, for certain ranges of the charge and spin Luttinger parameters, 
which have charge and spin conductances which are universal non-trivial numbers. 
A general solution for these fixed points remains an open problem. A simpler 
situation occurs in spin chains.\cite{Eggert} Again, I give only a telegraphic reminder of 
the field theory approach to the S=1/2 Heisenberg antiferromagnetic chain with Hamiltonian:
\be H=J\sum_j\vec S_j\cdot \vec S_{j+1}.\ee 
One method, is to start with the Hubbard model 
at 1/2-filling, where the charge excitations are gapped due to the Umklapp interaction. 
We may simply drop the charge boson from the low energy effective Hamiltonian 
which then contains only the spin boson or, equivalently, an $SU(2)_1$ WZW model. The 
low energy degrees of freedom of the spin operators occur at wave-vectors $0$ and $\pi$:
\be \vec S_j\approx {1\over 2\pi}[\vec J_L(j)+\vec J_R(j)]+(-1)^j\vec n(j),\ee
where the staggered component, of scaling dimension 1/2, can be written either in terms of a free boson, $\varphi$ and its 
dual $\theta$, with $g=1/2$ or else in terms of the primary field $g^\alpha_\beta$ of the WZW model:
\be \vec n\propto \hbox{tr}g\vec \sigma \propto \left(\begin{array}{c}
\cos (\varphi /\sqrt{2})\\
\sin  (\varphi /\sqrt{2})\\
\cos (\sqrt{2}\theta )
\end{array}\right).\label{n}\ee
The spin boson Hamiltonian contains the marginally irrelevant interaction, $-(g_s/2\pi )\vec J_L\cdot \vec J_R$, 
with a bare coupling constant $g_s$, of O(1). By including a second neighbour coupling, $J_2$, in the microscopic Hamiltonian the bare value of $g_s$ can be varied. At $J_2=J_{2c}\approx .2411J$, a phase transition 
occurs with the system going into a gapped spontaneously dimerized phase. In the low energy 
effective Hamiltonian the phase transition corresponds to the bare $g_s$ passing through zero 
and becoming marginally relevant rather than marginally irrelevant. 

A semi-infinite spin chain, $j\geq 0$ with a free BC corresponds to a D bc on $\theta$, just 
as for the fermionic model discussed above.\cite{Eggert} Then the staggered spin operator at zero becomes:
\be \vec n(0)\propto \left(\begin{array}{c}
\cos [\sqrt{2}\varphi_L(0)]\\
\sin  [\sqrt{2}\varphi_L(0)]\\
\partial_x\varphi_L(0)
\end{array}\right).\ee
All 3 components now have scaling dimension 1, and, it is easily seen, taking into account $SU(2)$ 
symmetry that:
\be \vec n(0)\propto \vec J_L(0).\ee
The D BC also implies $\vec J_L(0)=\vec J_R(0)$, so that both uniform and staggered spin 
components at $x=0$ reduce to $\vec J_L(0)$. Now consider the effect of a Kondo type coupling 
between a spin chain and one additional ``impurity spin''. In the simplest case where 
the impurity spin is also of size S=1/2, it makes an enormous difference exactly how 
it is coupled to the other spins. The simplest case where is where it is coupled at 
the end of a semi-infinite chain:
\be H=J'\vec S_1\cdot \vec S_2+J\sum_{i=2}^\infty \vec S_i\cdot \vec S_{i+1},\label{sckm}\ee
with the impurity coupling $J'\ll J$. For small $J'$ a low energy Hamiltonian 
description is appropriate, and since $\vec S_2\propto \vec J_L(0)$, we obtain 
the continuum limit of the Kondo model with a bare Kondo coupling $\lambda\propto J'$. \cite{Eggert,Laflorencie}
Thus we can take over immediately all the RG results on the Kondo effect except 
that we must beware of logarithmic corrections arising from the bulk marginal coupling constant, $g_s$, 
which are absent for the free fermion Kondo model. The correspondance with 
the free fermion Kondo model becomes nearly perfect when a bulk second neighbour 
interaction, $J_2$ is added to the Hamiltonian and fine-tuned to the critical 
point where this bulk marginal interaction vanishes.  Then only truly 
irrelevant bulk interactions (of dimension 4 or greater) distinguish the two models. 
The strong coupling fixed point of this Kondo model 
simply corresponds to the impurity spin being adsorbed into the chain, and corresponds 
to a renormalized $J'\to J$ at low energies. A more interesting model involves an 
impurity spin coupled to 2 semi-infinite chains:
\be H=J\sum_{-\infty}^{-2}\vec S_j\cdot \vec S_{j+1}+J\sum_{j=1}^\infty \vec S_j\cdot \vec S_{j+1}
+J'\vec S_0\cdot (\vec S_{-1}+\vec S_1).\ee
The continuum limit is now the 2-channel Kondo model,\cite{Eggert} with the left and right sides of the impurity 
corresponding to the 2 channels. Again the Kondo fixed point simply corresponds to a ``healed chain''
with $J'$ renormalizing to $J$ and a restoration of translational invariance at low energies. 
Other possibilties involve a ``side-coupled'' impurity spin. For example we may couple the 
impurity spin $\vec S'$ to one site on a uniform chain:
\be H=J\sum_{j=-\infty}^\infty \vec S_j\cdot \vec S_{j+1}+J'\vec S'\cdot \vec S_0.\ee
Now the correspondance to the ordinary free fermion Kondo model fails dramatically 
because the boundary interaction $\propto J'\vec S'\cdot \vec n(0)$ appears in the 
effective Hamiltonian where $\vec n$ is the staggered spin operator introduced in Eq. (\ref{n}). 
This is a strongly relevant dimension 1/2 boundary interaction. It renormalizes to 
infinity.  It is easy to understand the low energy fixed point in this case by 
imagining an infinite bare $J'$.  The the impurity spin forms a singlet with $\vec S_0$ 
and the left and right sides of the chain are decoupled. The stability of such a fixed 
point is verifed by the fact that the spins at the end of the open chains, $\vec S_{\pm 1}$ 
have dimension 1 so that an induced weak ``bridging'' coupling $J_{eff}\vec S_{-1}\cdot \vec S_1$ 
has dimension 2 and is thus an irrelevant boundary interaction. We expect even a small $J'$
to renormalize to such a strong coupling fixed point but in general the screening 
cloud will be spread over longer distances.  Nonetheless, the left and right sides 
decouple at low energies and long distances.  Other examples, including larger spin impurities, 
were discussed in [\onlinecite{Eggert}].  

We may also couple an impurity spin to a Hubbard type model with gapless spin {\it and} charge 
degrees of freedom. The various situations closely parallel the spin chain case. In particular 
the cases of the impurity spin at the end of the chain or embedded in the middle still 
correspond essentially to the simple Kondo model. This follows because the D BC on 
both spin and charge bosons has the effect of reducing both uniform {\it and staggered} 
spin density operators at the boundary to $\vec J_L(0)$.  This model, which can 
be applied to a quantum dot coupled to a quantum wire in a semi-conductor heterostructure, 
was analysed in detail in [\onlinecite{Pereira}].

\section{Quantum Impurity Entanglement Entropy}
Quantum entanglement entropy has become a popular subject in recent years because of 
its connection with black holes, quantum computing and the efficiency of 
the Density Matrix Renormalization Group method, and its generalizations, 
for calculating many body groundstates (on a classical computer). In this 
lecture I will discuss the intersection of this subject with quantum impurity 
physics.\cite{Sorensen} After some generalities, I will focus on the simple example 
of the single channel Kondo model, obtaining a novel perspective on 
the nature of the Kondo groundstate and the meaning of the characteristic 
length scale $\xi_K$. In the second lecture,  I discussed and defined the zero temperature impurity entropy, 
showing that it is a universal quantity, characterising the BCFT fixed point, 
and always decreasing under boundary RG flow. Quantum entanglement 
entropy is, in general, quite distinct from thermodynamic entropy, 
being a property of a quantum ground state and depending on an arbitrary  
division of a system into two different spatial regions. Nonetheless, as we shall 
see the thermodynamic impurity entropy, in the $T=0$ limit, also appears as a 
term in the entanglement entropy, in a certain limit. 

Consider first a CFT with central charge $c$ on a semi-infinite interval, $x>0$, with a CIBC, labelled $A$, 
at $x=0$. We trace out the region, $x'\geq x$ to define the density matrix, 
and hence the entanglement entropy, $S_A(x)=-\hbox{tr}\rho\ln \rho $, for the region, $0\leq x'\leq x$. 
[Note that I am using the natural logarithm in my definition of entanglement entropy.  Some authors 
define $S$ using the logarithm base 2, which simply divides $S$ by $\ln 2$.] 
Calabrese and Cardy (C\&C) showed,\cite{Calabrese} generalizing earlier results of 
Holzhey, Larsen and Wilczek \cite{Holzhey}, that 
this entangement entropy is given by:
\be S_A(x)=(c/6) \ln (x/a) +c_A.\label{Sent0}\ee
Here $a$ is a non-universal constant.  $c_A$ is another constant which could have 
been adsorbed into a redefinition of $a$. However, $S_A(x)$ is written this way because, by construction,  
the constant $a$ is independent of the choice of CIBC, $A$, while the constant $c_A$ depends on it. 
C\&C showed that the generalization of $S_A(x)$ to a finite inverse temperature, $\beta$, is given by a 
standard conformal transformation:
\be S_A(x,\beta )=(c/6)\ln [(\beta /\pi a)\sinh (2\pi x/\beta )]+c_A.\ee
$S_A(x,\beta )$ is defined by beginning with the Gibbs density matrix for the entire system, $e^{-\beta H}$ 
and then again tracing out the region $x'>x$. 
Now consider the high temperatures, long length limit, $\beta \ll x$:
\be S_A\to 2\pi cx/\beta +(c/6)\ln (\beta /2\pi a)+c_A+O(e^{-4\pi x/\beta}).\label{Sas}\ee
The first term is the extensive term (proportional to $x$) in the thermodynamic entropy for 
the region, $0<x'<x$. 
The reason that we recover the thermodynamic entropy when $x\gg \beta$ is because, 
in this limit, we may regard the region $x'>x$ as an ``additional reservoir'' for the 
region $0\leq x'\leq x$. That is, the thermal density matrix can be defined by integrating 
out degrees of freedom in a thermal reservoir, which is weakly coupled to the entire system. 
On the other hand, the region $x'>x$ is quite strongly coupled to the region $x'<x$. 
Although this coupling is quite strong, it only occurs at one point, $x$. When 
$x\gg \beta$, this coupling only weakly perturbs the density matrix for the region $x'<x$. 
Only low energy states, with energies of order $1/x$ and a neglible fraction of the 
higher energy states (those localized near $x'=x$) are affected by the coupling 
to the region $x'>x$. The thermal entropy for the system, with the boundary at $x=0$ in the limit $x\gg \beta$ is:
\be S_{A,th}\to 2\pi cx/\beta + \ln g_A+\hbox{constant},  \ee
with corrections that are exponentially small in $x/\beta$.
The only dependence on the CIBC, in this limit, is through the constant term, $\ln g_A$, the impurity entropy. 
Thus it is natural to identify the BC dependent term in the entanglement entropy 
with the BC dependent term in the thermodynamic entropy:
\be c_A=\ln g_A.\ee
This follows since, in the limit, $x\gg \beta$, we don't expect the coupling to the 
region $x'>x$ to affect the thermodynamic entropy associated with the boundary $x'=0$, $c_A$. 
Note that the entanglement entropy, Eq. (\ref{Sas}), contains an additional large term 
not present in the thermal entropy. We may ascribe this term to a residual effect of the 
strong coupling to the region $x'>x$ on the reduced density matrix. However, this extra term does not depend on the CIBC 
as we would expect in the limit $x\gg\beta$ in which the ``additional reservoir'' is far from the boundary. 
Now passing to the opposite limit $\beta \to \infty$, we obtain the remarkable result that 
the only term in the (zero temperature) entanglement entropy depending on the BC is precisely 
the impurity entropy, $\ln g_A$.  

Since the impurity entropy, $\ln g_A$, is believed to 
be a universal quantity characterizing boundary RG fixed points, it follows that the 
boundary dependent part of the ($T=0$) entanglement entropy also enjoys this property. 
In particular, we might then expect this quantity to exhibit an RG crossover as 
we increase $x$.  That is, consider the entanglement entropy, $S(x)$, for the 
type of quantum impurity model discussed in these lectures, 
that is described by a conformal field theory in the bulk (at low energies) and has 
more or less arbitrary boundary interactions. As we increase $x$, we might 
expect $S(x)$ to approach the CFT value, Eq. (\ref{Sent0}), with the 
value of $c_A$ corresponding to the corresponding CIBC. More interestingly, 
consider such a system which is flowing between an unstable and a stable 
CIBC, $A$ and $B$ respectively, such as a general Kondo model with weak bare couplings. Then as discussed 
in lectures 1 and 2, the impurity part of the thermodynamic entropy  
crosses over between two values $\ln g_A$ and $\ln g_B$. In the general $k$-channel Kondo 
case, $g_A=(2S+1)$, the degeneracy of the decoupled impurity spin. $g_B$ is determined 
by the Kondo fixed point, having the values:
\bea g_B&=&2S'+1,\ \ (S'\equiv S-2k, k\leq 2S)\nonumber \\
&=&{\sin [\pi (2s+1)/(2+k)]\over \sin [\pi /(2+k)]},\ \  (k>2S).\eea
This {\it change} in $\ln g_A$ ought to be measureable in numerical simulations 
or possibly even experiments.  

To make this discussion more precise, it is convenient 
to define a ``quantum impurity entanglement entropy'' as the difference 
between entanglement entropies with and without the impurity. Such a 
definition parallels the definition of impurity thermodynamic entropy 
(and impurity susceptibility, impurity resistivity, etc.) used in lectures 1 and 2. 
To keep things simple, consider the Kondo model in 3 dimensions and consider 
the entanglement of a spherical region of radius $r$ containing the impurity 
at its centre, 
with the rest of the system (which we take to be of infinite size, for now). 
This entanglement entropy could be measured both before and after adding 
the impurity, the difference giving the impurity part. Because of the 
spherical symmetry it is not hard to show that the entanglement entropy 
reduces to a sum of contributions from each angular momentum channel, $(l,m)$. 
For a $\delta$-function Kondo interaction, only the s-wave harmonic is 
affected by the interaction.  Therefore, the impurity entanglement entropy 
is determined entirely by the s-wave harmonic and can thus be 
calculated in the usual 1D model. Thus we may equivalently consider 
the entanglement between a section of the chain, $0<r'<r$, including the impurity, 
with the rest of the chain. It is even more convenient, especially for 
numerical simultations, to use the equivalent model, discussed in the previous lecture, 
 of an impurity spin  weakly coupled at the end of a Heisenberg S=1/2  spin chain, 
Eq. (\ref{sckm}). Region $A$ is the first $r$ sites of the chain, which in general 
has a finite total length $R$. The corresponding entanglement entropy is written 
as $S(J_K',r,R)$ where we set $J=1$ and now refer to the impurity (Kondo) coupling 
as $J_K'$. The long distance behavior of $S(r)$ is found to have both uniform and 
alternating parts:
\be S(r)=S_U(r)+(-1)^rS_A(r),\ee
where both $S_U$ and $S_A$ are slowly varying. I will just focus here on $S_U$ 
which we expect to have the same universal behaviour as in other realizations 
of the Kondo model, including the 3D free fermion one. We define the impurity 
part of $S$ precisely as:
\be S_{imp}(J_K',r,R)\equiv S_U(J_K',r,R)-S_U(1,r-1,R-1),\ \  (r>1).\label{S_U}\ee
Note that we are subtracting the entanglement entropy when the first spin, at site $j=1$, 
is removed.  This removal leaves a spin chain of length $R-1$ with all couplings equal to 1. 
After the removal, region $A$ contains only $r-1$ sites. If we start with a weak Kondo coupling, $J'\ll J=1$, 
we might expect to see cross over between weak and strong coupling fixed points 
as $r$ is increased past the Kondo screening cloud size $\xi_K$. Ultimately, this behaviour 
was found but with a surprising dependence on whether $R$ is even or odd. [Note that 
is a separate effect from the dependence on whether $r$ is even or odd which I have already 
removed by focussing on the uniform part, defined in Eq. (\ref{S_U}).]

As usual, I first focus on the behaviour near the fixed points, where perturbative RG methods can be used. 
Strong coupling perturbation theory in the LIO at the Kondo fixed point turns 
out to be very simple. The key simplifying feature is that the Fermi liquid, LIO is 
proportional to the energy density itself, at $r=0$. It follows from Eq. (\ref{Heff}) and 
(\ref{Sug}) (in the single channel $k=1$ case) that the low energy effective Hamiltonian 
including the LIO at the Kondo fixed point can be written, in the purely left moving formalism:
\be H={1\over 6\pi}\int_{-R}^R dx\vec J_L(x)^2-{\xi_K\over 6}\vec J_L^2(0)
=\int_{-R}^R dx{\cal H}(x) -\pi \xi_K{\cal H}(0),\ee
where ${\cal H}(x)\equiv (1/6\pi )\vec J_L^2(x)$ is the energy density. (I have 
set $v=1$ and used $\xi_K=1/T_K$.) 
The method of Holzhey-Wlczek and Calabrese-Cardy for calculating the entanglement entropy 
is based on the `` replica trick''. That is to say, the partition function, $Z_n$ 
is calculated on an $n$-sheeted Reimann surface, ${\cal R}_n$, with the sheets joined along region $A$, 
from $r'=0$ to $r'=r$. The trace of the  $n^{th}$ power of the reduced density matrix 
can be expressed as:
\be \hbox{Tr}\rho (r)^n={Z_n(r)\over Z^n}\label{Z_n}\ee
where $Z$ is the partition function on the normal complex plane, ${\cal C}$. The entanglement entropy 
is obtained from the formal analytic continuation in $n$:
\be S=-\lim_{n\to 1}{d\over dn}[\hbox{Tr}\rho (r)^n].\label{repl}\ee
The correction to $Z_n(r)$ of first order in $\xi_K$ is:
\be \delta Z_n=(\xi_K\pi )n\int_{-\infty}^\infty d\tau <{\cal H}(\tau ,0)>_{{\cal R}_n}.\label{Zncorr}\ee
C\&C showed related this expectation value of the energy density on ${\cal R}_n$ to 
a 3-point correlation function on the ordinary complex plane:
\be <{\cal H}(\tau ,0)>_{{\cal R}_n}={<{\cal H}(\tau ,0)\varphi_n(r)\varphi_{-n}(-r)>_{\cal C}
\over <\varphi_n(r)\varphi_{-n}(-r)>_{\cal C}}.\label{HRn}\ee
Here the primary operators $\varphi_{\pm n}$ sit at the branch points $\pm r$ (in the 
purely left moving formulation) and have scaling dimension
\be \Delta_n=(c/24)[1-(1/n)^2]\ee
where, in this case, the central charge is $c=1$. Thus Eq. (\ref{HRn}) gives:
\be <{\cal H}(\tau ,0)>_{{\cal R}_n}={[1-(1/n)^2]\over 48\pi}{(2ir)^2\over (\tau -ir)^2(\tau +ir)^2}.\ee
Doing the $\tau$-integral in Eq. (\ref{Zncorr}) gives:
\be \delta Z_n=-{\xi_K\pi \over 24r}n[1-(1/n)^2].\ee
Since there is no correction to $Z$, to first order in $\xi_K$, inserting this result in 
Eqs. (\ref{Z_n}) and (\ref{repl}) gives:
\be S_{imp}={\pi \xi_K\over 12r}.\ee
Here we have used the fact that $g_A=1$, $c_A=0$ at the Kondo fixed point.
C\&C also observed that the entanglement entropy for a finite total system size $R$, can 
be obtained by a conformal transformation.  For a conformally invariant system this generalizes 
Eq. (\ref{Sent0}) to:
\be S(r,R)=(c/6)\ln [(R/\pi a)\sin (\pi r/R)]+c_A.\ee
Our result for $S_{imp}$ can also be extended to finite $R$ by the same conformal transformation of 
the 2-point and 3-point functions in Eq. (\ref{HRn}).  The 3-point function now 
has a disconnected part, but this is cancelled by the correction to $Z^n$ of first order in $\xi_K$, 
leaving:
\be {\delta Z^n\over Z^n}={\xi_K\pi n[1-(1/n)^2]\over 48\pi}\int_{-\infty}^\infty d\tau 
\left[{(\pi /R)\sinh 2i\pi r/R\over \sinh [\pi (\tau +ir)/R]\sinh [\pi (\tau -ir)/R]}\right]^2.\ee
Doing the integral and taking the replica limit now gives:
\be S_{imp}(r,R)={\pi \xi_K\over 12R}\left[1+\pi \left(1-{r\over R}\right)\cot \left(\pi r\over R\right)\right].
\label{SimpFLT}\ee
We emphasize that these results can only be valid at long distances where we can use FLT, 
i.e. $r\gg \xi_K$. They represent the first term in an expansion in $\xi_K/r$. 

The thermodynamic impurity entropy is known to be a universal scaling function of $T/T_K$. It seems 
reasonable to hypothesize that the impurity entanglement entropy (for infinite system size) is 
a universal scaling function of $r/\xi_K$. At finite $R$, we might then expect it to be a 
universal scaling functions of the two variables $r/\xi_K$ and $r/R$. Our numerical results 
bear out this expectation with one perhaps surprising feature: While the scaling function is 
independent of the total size of the system at $R\to \infty$ there is a large difference 
between integer and half-integer total spin (i.e. even and odd $R$ in the spin chain version of the 
Kondo model) for finite $R$;  i.e. we must define two universal scaling functions 
$S_{imp,e}(r/\xi_K,r/R)$ and $S_{imp,o}(r/\xi_K,r/R)$ for even and odd $R$.  These become 
the same at $r/R\to \infty$. 

\begin{figure}[th]
\begin{center}
\includegraphics[clip,width=8cm]{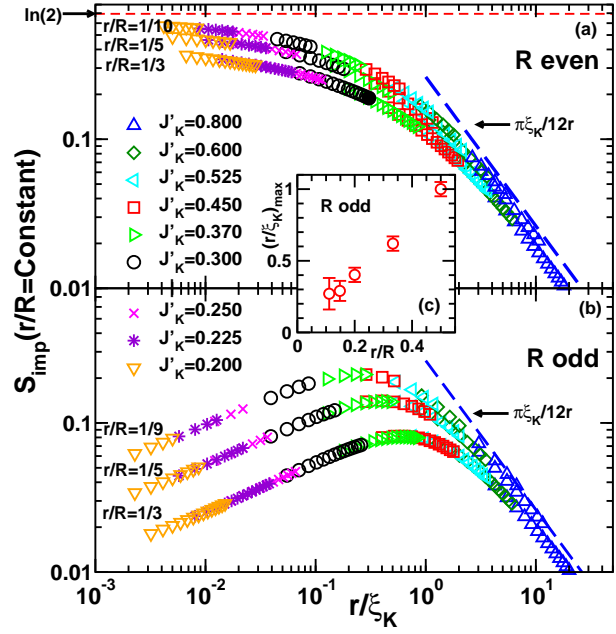}
\caption{Impurity entanglement entropy for fixed $r/R$ plotted versus $r/\xi_K$ for 
both cases $R$ even and $R$ odd. FLT predictions for large $r/\xi_K$ are shown.  Inset: 
location of the maximum, $(r/\xi_K)_{\hbox{max}}$ for odd $R$ plotted versus $r/R$.}
\label{fig:Sent}
\end{center}
\end{figure}

\begin{figure}[th]
\begin{center}
\includegraphics[clip,width=8cm]{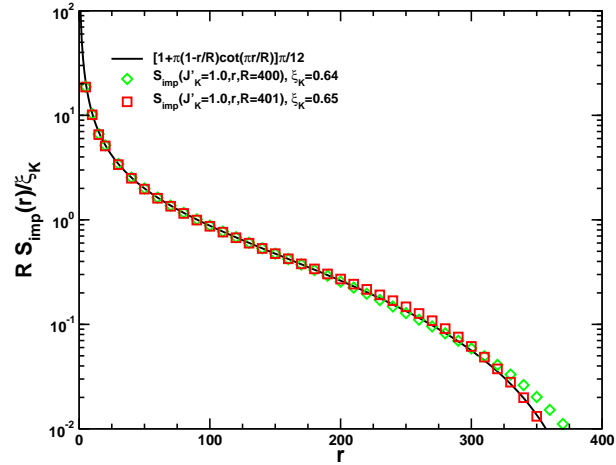}
\caption{Impurity entanglement entropy for the Kondo spin chain model calculated 
by DRMG compared with the Fermi liquid calculation of Eq. (\ref{SimpFLT}).}
\label{fig:SentFL}
\end{center}
\end{figure}

We calculated the impurity entanglement entropy numerically using the Density Matrix Renormalization 
Group (DMRG) method.  In this approach a chain system is built up by adding pairs of additional sites 
near the centre of the chain and systematically truncating the Hilbert Space at a manageable size 
(typically around 1,000) at each step. The key feature of the method is the choice of which 
states to keep during the truncation. It has been proven that the optimium choice is the 
eigenstates of the reduced density matrix with the largest eigenvalues. Since the method, by 
construction, calculates eigenvalues of $\rho_A$, it is straightforward to calculate 
the corresponding entanglement entropy. Some of our results for $S_{imp}(r,R,J_K')$ are 
shown in Figs. (\ref{fig:Sent}) and (\ref{fig:SentFL}). The second figure shows that the 
large $r$ result of Eq. (\ref{SimpFLT})  works very well.  This is rather remarkable confirmation of the 
universality of the quantum impurity entanglement entropy since we have obtained results for the microscopic model 
using a continuum field theory. It supports the idea that the impurity entanglement entropy 
would be given by the same universal functions for other realizations of the Kondo model 
including the standard 3D free fermion one. Note that the even and odd scaling functions 
have very different behaviour when $R<\xi_K$. For the integer spin case, $S_{imp}$ increases 
monotonically with decreasing $r$ and appears to 
be approaching $\ln 2$ in the limit $r\ll \xi_K$, $R$. This is what is expected from the 
general C\&C result since $g_A=\ln 2$ at the weak coupling fixed point with a decoupled 
impurity spin. On the other hand, for the half-integer spin case $S_{imp}$ initially 
increases with decreasing $r/\xi_K$ but eventually goes through a maximum and starts 
to decrease. The maximum occurs when $r/\xi_K\approx r/R$; i.e. when $\xi_K\approx R$. 

These results can be understood heuristically from a resonating valence bond picture 
of the ground spin in the Kondo spin chain model. Let's first consider the case of 
integer total spin, $R$ even. Then the ground state is a spin singlet. It is important 
to realize that when $J_K'\to 0$ the singlet ground state becomes degenerate with a triplet 
state.  This occurs since the ground state of the other $R-1$ sites, $j=2,3,\ldots R$ has
spin-1/2, as does the impurity spin and the two are coupled. However, by continuity in $J_K'$, 
it is the singlet state which is considered. This singlet state has strong entanglement 
of the impurity spin with the rest of the system, despite the fact that there is 
no term in the Hamiltonian coupling them together when $J_K'=0$.
Any singlet state, and hence the ground state for any value of $J_K'$, can be 
written as some linear combination of ``valence bond states'' i.e. states in which 
pairs of spins form a singlet. (We can always restrict to ``non-crossing'' states 
such that if we draw lines connecting every pair of contracted spins, none of 
these lines cross each other.) If we consider a ``frozen'' valence bond state, 
i.e. any particular basis state, then the entanglement entropy is simply $\ln 2$ 
times the number of valence bonds from region $A$ to $B$. 
There will always be a valence bond connecting  
the impurity spin to some other spin in the system; we refer to this as the 
Impurity Valence Bond (IVB). Intuitively, if the IVB connects the impurity 
to a spin outside of region A then we think of this as resulting in an impurity entanglement of $\ln 2$; however this picture is certainly naive 
because the valence bond basis, while complete, is not orthogonal.  
 We may think 
of the typical length of the IVB as being $\xi_K$ since the spin screening the impurity 
is precisely the one forming the IVB. This picture makes it quite clear why $S_{imp}$ is 
a decreasing function of $r$ and why $\xi_K$ is the characteristic scale for its variation. 

Now consider the case of half-integer spin, $R$ odd. In this case, when $J_K'=0$ there 
is zero entanglement between the impurity spin and the rest of the system. In this 
case the other $R-1$ sites have a spin zero ground state, decoupled from the impurity, 
which is unpaired. For $R$ odd and any $J_K'$ the ground state always has spin 1/2.  This 
can again be represented as an RVB state but each basis valence bond state has precisely 
one unpaired spin, which may or may not be the impurity. At $J_K'=0$ is {\it is} the impurity 
with probability 1, but consider what happens as we increase $J_K'$ from zero, corresponding to decreasing 
$\xi_K\propto \exp [\hbox{constant}/J_K']$ from infinity. The probability of having an IVB 
increases.  On the other hand, the typical length of the IVB when it is present is $\xi_K$ 
which decreases. These two effects trade off to give a peak in $S_{imp}$ when 
$\xi_K$ is approximately $R$. 

Our most important conclusion is probably that quantum impurity entanglement entropy appears 
to exihibit universal cross over between boundary RG fixed points with the size, $r$, of 
region $A$ acting as an infrared cut-off.

\section{Y-junctions of quantum wires}
\begin{figure}[th]
\begin{center}
\includegraphics[clip,width=8cm]{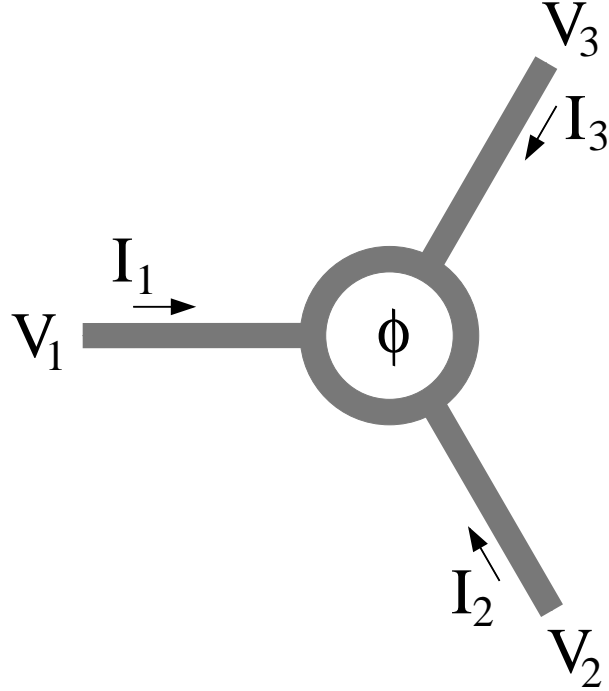}
\caption{A Y-junction with voltages and currents indicated.}
\label{fig:Yjunction}
\end{center}
\end{figure}

Now I consider 3 semi-infinite spinless  Luttinger liquid quantum wires, 
all with the same Luttinger parameter, $g$, meeting at 
a Y-junction,\cite{Oshikawa} as shown in Fig. (\ref{fig:Yjunction}).
  By imposing a magnetic field near the junction we can introduce a non-trivial phase, $\phi$ 
into the tunnelling terms between the 3 wires. A corresponding lattice model is:
\be H=\sum_{n=0}^\infty\sum_{j=1}^3[-t(\psi^\dagger_{n,j}\psi_{n+1,j}+h.c.)+\tilde V\hat n_{n,j}\hat n_{n+1,j}]
-(\tilde \Gamma /2)\sum_{j=1}^3[e^{i\phi /3}\psi^\dagger_{0,j}\psi_{0,j-1}+h.c.].\label{YHam}\ee
(In general we can also introduce potential scattering terms at the end of each wire.) 
Here $j=1$, $2$, $3$ labels the 3 chains cylically so that we identify $j=3$ with $j=0$. 
We bosonize, initially introducing a boson $\varphi_j(x)$ for each wire: 
\be \psi_{j,L/R}\propto \exp [i(\varphi_j\pm \theta_j)/\sqrt{2}].\ee
It follows immediately from the discussion of the 2-wire case in the 4$^{th}$ lecture that 
$\tilde \Gamma$ is irrelevant for $g<1$, the case of repulsive interactions. Thus 
we expect the decoupled wire, zero conductance fixed point to be the stable one in that case. 
On the other hand, if $g>1$ the behaviour is considerably more complex and interesting. This case corresponds 
to effectively attractive interaction between electrons, as can arise from phonon exchange. 

It is convenient to make a basis change, analgous to the even and odd channel introduce in Lecture 4 for 
the 2 wire junction:
\bea \Phi_0&=&{1\over \sqrt{3}}(\varphi_1+\varphi_2+\varphi_3)\nonumber \\
\Phi_1&=&{1\over \sqrt{2}}(\varphi_1-\varphi_2)\nonumber \\
\Phi_2&=&{1\over \sqrt{6}}(\varphi_1+\varphi_2-2\varphi_3)\label{Phi}\eea
and similarly for the $\theta_i$'s. It is also convenient to introduce 3 unit vectors, 
at angles $2\pi /3$ with respect to each other, 
acting on the $(\Phi_1,\Phi_2)$ space:
\bea \vec K_1&=&(-1/2,\sqrt{3}/2)\nonumber \\
\vec K_2&=&(-1/2,-\sqrt{3}/2)\nonumber \\
\vec K_3&=&(1,0).\eea
We will also use 3 other unit vectors rotated by $\pi /2$ relative to these ones 
which we write as $\hat z \times \vec K_i=(-K_{iy},K_{ix})$.
The various boundary interactions are now written in this basis:
\bea 
T_{21}^{RL}&=&\psi_{2R}^\dagger\psi_{1L}\propto e^{i\vec K_3\cdot \vec \Phi}e^{i(1/\sqrt{3})\hat z \times \vec K_3
\cdot \vec \Theta}e^{i\sqrt{2/3}\Theta_0}\nonumber \\
T_{12}^{RL}&=&\psi_{1R}^\dagger\psi_{2L}\propto e^{-i\vec K_3\cdot \vec \Phi}e^{i(1/\sqrt{3})\hat z \times \vec K_3
\cdot \vec \Theta}e^{i\sqrt{2/3}\Theta_0}\nonumber \\
T_{21}^{LL}&=&\psi_{2L}^\dagger\psi_{1L}\propto e^{i\vec K_3\cdot \vec \Phi}e^{i\vec K_3
\cdot \vec \Theta}\nonumber \\
T_{11}^{RL}&=&\psi^\dagger_{R1}\psi_{L1}\propto e^{-i(2/\sqrt{3})\hat z\times \vec K_1\cdot \Theta}
e^{i\sqrt{2/3}\Theta_0}
\label{PT}\eea
et cetera. Note that we have not yet imposed any particular BC's on the fields at the origin. A more 
standard approach would probably be to start with the $\tilde \Gamma =0$ BC, $\Theta_i(0)$= constant 
and then allow for the possibility that these BC's renormalize due to the tunnelling. We call 
the current approach the method of ``Delayed Evaluation of Boundary Conditions''. 
 Hower, we expect the ``centre of mass'' field, $\theta_0$ is always pinned, $\Theta_0 (0)=$ constant. 
Since $\Phi_0$ carries a non-zero total charge, and hence doesn't appear in any of the boundary interactions, 
it would not make sense for any other type of boundary condition to occur in the ``0'' sector.  We 
may thus simply drop the factor involving $\Theta_0$ from all the boundary interactions; it 
makes no contribution to scaling dimensions. 
However, we must consider other possible BC's in the $1$, $2$ sector, since, for $g>1$ the 
simple D BC on $\vec \Theta$ is not stable, as mentioned above. Another simple possibility 
would be a D BC on $\vec \Phi$: $\Phi_i(0)=c_i$ for two constants $c_i$. To check the 
stability of this BC under the RG we must consider the LIO. The various candidates 
are the tunnelling and backscattering terms in Eq. (\ref{PT}).  Imposing the BC $\vec \Phi =$ constant, 
we can evaluate the scaling dimension of the exponential factors involving $\vec \Theta (0)$ 
by the same method used in Lecture 4. The BC implies that we should replace 
$\vec \Theta (0)$ by $2\sqrt{g}\vec \Phi_L(0)$. The dimensions of these operators can 
then be read off:
\bea \Delta_{21}^{RL}&=&g/3\nonumber \\
\Delta_{21}^{LL}&=&g\nonumber \\
\Delta_{11}^{RL}&=&4g/3,\eea
et cetera. $T_{ij}^{RL}$, for $i\neq j$ are the LIO's. 
 They are relevant for $g<3$. Thus we see that this cannot be the stable fixed point for $1<g<3$. 

Stable fixed points, for $1<g<3$ can be identified from the form of the $T^{RL}_{j,j\pm 1}$. 
If $\tilde \Gamma$ grows large under renormalization it is plausible that one or 
the other of this set of operators could develop an expectation value. Note that if:
\be <T_{j,j+1}>\neq 0,\ee
this would correspond to strong tunnelling from $j$ to $j+1$. On the other hand, if:
\be <T_{j,j-1}>\neq 0,\ee
this would correspond to strong tunnelling from $j$ to $j-1$. Breaking time reversal 
by adding a non-zero magnetic flux, $\phi$, favours one or the other of these 
tunnelling paths, dependng on the sign of $\phi$. On the other hand, in the time-reversal 
invariant case $\phi =0$ (or $\pi$) we do not expect such as expectation value to develop. 
These fixed points obey the BC's:
\be \pm \vec K_i\cdot \vec \Phi (0)+\sqrt{1/3}(\hat z\times \vec K_i)\cdot \vec \Theta (0)=\vec C,\ \  (i=1,2,3),
\label{cfp}\ee
for some constants $C_i$. 
Note that since $\sum_{i=1}^3\vec K_i=0$, these are only 2 independent BC's. Introducing left and right 
moving fields:
\bea \vec \Phi &=& {1\over \sqrt{g}}(\vec \Phi_L+\vec \Phi_R)\nonumber \\
\vec \Theta &=&\sqrt{g}(\vec \Phi_L-\vec \Phi_R),\eea
we may write these BC's as:
\be \vec \Phi_R(0)={\cal R}\vec \Phi_L(0)+\vec C',\label{bcRL}\ee
where $\vec C'$ is another constant vector and ${\cal R}$ is a orthogonal matrix which 
we parameterize as:
\be {\cal R} = \left(\begin{array}{rr}
\cos \xi & -\sin \xi \\
\sin \xi & \cos \xi
\end{array}\right),\label{Rxi}\ee
with
\be \xi = \pm 2\arctan (\sqrt{3}/g).\label{xi}\ee
We refer to these as the ``chiral'' fixed points, $\chi_{\pm}$, with the $+$ or $-$ corresponding 
to the sign in Eqs. (\ref{cfp}) and (\ref{xi}). Note that if $\xi=0$ we obtain the usual D BC on $\vec \Theta$ 
and if $\xi =\pi$ we obtain a D BC on $\vec \Phi$. The chiral BC's are, in a sense, intermediate 
between these other two BC's. 

As usual, we check their stability 
by calculating the scaling dimension of the LIO. Once we have the BC's in the form of Eq. (\ref{bcRL}) 
it is straightforward to calculate the scaling dimension of any vertex operators of the general form:
\be {\cal O}=\exp \left( i\sqrt{g}\vec a\cdot \vec \Phi + i{1\over \sqrt{g}}\vec b\cdot \vec \Theta \right) 
=\exp [i(\vec a-\vec b)\cdot \vec \Phi_R+i(\vec a +\vec b)\cdot \vec \Phi_L]
\propto \exp \{i[{\cal R}^{-1}(\vec a-\vec b)+(\vec a + \vec b)]\cdot \vec \Phi_L\}
.\ee
\be \Delta_{\cal O}={1\over 4}|{\cal R}(\vec a + \vec b)+(\vec a - \vec b)|^2.\ee
Applying this formula to the $\chi_{\pm}$ fixed points we find that 
all the operators listed in Eq. (\ref{PT}) have the same dimension:
\be \Delta = {4g\over 3+g^2}.\ee
Since $\Delta >1$ for $1<g<3$, we conclude that the chiral fixed points are stable for this 
intermediate range of $g$, only. 
Thus we hypothesize that the system renormalizes to these chiral fixed points for this range of 
$g$ whenever there is a non-zero flux, $\phi \neq 0$. However, these fixed points are 
presumably not allowed due to time reversal symmetry when $\phi =0$ and there 
must be some other stable fixed point to which the system renormalizes. This fixed 
point, which we referred to as ``M'' appears to be of a less trivial type than 
there ``rotated D'' states. An attractive possibility is that the M fixed point 
is destablized by an infinitesimal flux leading to a flow to one of the chiral 
fixed points.  Alternatively, it is possible that there is a critial value of the flux, $|\phi_c|\neq 0$ 
neccessary to destabilize the M fixed point. In that case there would presumably be two more, as yet undetermined,  
CI BC's corresponding to these critical points.

\subsection{Conductance}
Once we have identified the CIBC's it is fairly straightforward to calculate the conductance 
using the Kubo formula. For a Y-junction the conductance is a tensor.  If we apply voltages $V_i$ 
far from the junction on lead $i$ and let $I_i$ be the current, flowing towards the junction, 
on lead $i$, then, for small $V_i$, 
\be I_i=\sum_{j=1}^3G_{ij}V_j.\ee
Since there must be no current when all the $V_i$ are equal to each other and since the 
total current flowing into the junction is always zero, it follows that 
\be \sum_iG_{ij}=\sum_jG_{ij}=0.\ee
Also taking into account the $Z_3$ symmetry of the model, we see that the most 
general possible form of $G_{ij}$ is:
\be G_{ij}={G_S\over 2}(3\delta_{ij}-1)+{G_A\over 2}\epsilon_{ij}.\label{GSA}\ee
Here $\epsilon_{ij}$ is the $3\times 3$ anti-symmetric, $Z_3$-symmetric tensor with $\epsilon_{12}=1$. 
$G_A$, which is odd under time reversal, can only be non-zero when there is a non-zero flux. 

In the non-interacting case, $\tilde V=0$ in Eq. (\ref{YHam}), $g=1$, we may calculate 
the conductance by a simple generalization of the Landauer formalism. We imagine 
that the three leads are connected to distant reservoirs at chemical potentials 
$\mu -eV_j$. Each reservoir is assumed to emit a thermal distribution of 
electrons down the lead and to perfectly adsorb electrons heading towards it.  
The conductance can then be expressed in terms of the $S$-matrix for the Y-junction.
This is defined by solutions of the lattice Schroedinger equation:
\bea -t(\Phi_{n+1,j}+\Phi_{n-1,j})&=&E\Phi_{n,j},\ \  (n\geq 1)\nonumber \\
-t\Phi_{1,j}-(\tilde \Gamma /2)(e^{i\phi /3}\Phi_{0,j-1}+e^{-i\phi /3}\Phi_{0,j+1})&=&E\phi_{o,j}.\eea
The wave-functions are of the form:
\be\Phi_{n,j}A_{\hbox{in},j}e^{-ikn}+A_{\hbox{out},j}e^{ikn},\ee
for all $n\geq 0$ and $j=1$, $2$, $3$, with energy eigenvalues $E=-2t\cos k$. 
The $3\times 3$ S-matrix is defined by:
\be \vec A_{\hbox{out}}=S\vec A_{\hbox{in}}.\ee
The most general $Z_3$ symmetric form is:
\bea S_{ij}&=&S_0\ \  (i=j)\nonumber \\
&=& S_-\ \  (i=j-1)\nonumber \\
&=& S_+\ \  (i=j+1).\eea

In the time-reversal invariant case $S_+=S_-$.  It is then easy to see that 
unitarity of $S$ implies $|S_{\pm}|\leq 2/3$. In this case, for any wave-vector, $k$, 
$|S_{\pm}|$ reaches $2/3$ for some value of $\tilde \Gamma$ of O(1). It can also be checked 
that when $k-\phi = \pi$ and $\tilde \Gamma =2$, $S_+=0=S_0=0$ and $|S_-|=1$. In this case an electron 
incident on lead  $j$ is transmitted with unit probability to lead $j-1$. 
To calculate the Landauer conductance, we observe that the total current on lead $j$ 
is the current emitted by reservoir $j$, minus the reflected current plus the 
current transmitted from leads $j\pm 1$:
\be I_j=e\int {dk\over 2\pi}v(k)[(1-|S_{jj}|^2)n_F(\epsilon_k-eV_j)-\sum_{\pm}|S_{j,j\pm 1}|^2
n_F(\epsilon_k-eV_{j\pm 1})].\ee
 This gives the conductance tensor of the form of Eq. (\ref{GSA}) with:
\be G_{S/A}={e^2\over h}(|S_+|^2\pm |S_-|^2),\ee
where $S_{\pm}$ are now evaluated at the Fermi energy. 
Thus the maximum possible value of $G_S$,in the zero flux 
case, is $(8/9)e^2/h$, 
 when $|S_{\pm}|=2/3$. On the other hand, for non-zero flux, when $S_+=S_0=0$, $G_S=-G_A=e^2/h$, 
i.e. $G_{ii}=-G_{i,i+1}=e^2/h$ but $G_{i,i-1}=0$. This implies that if a voltage 
is imposed on lead $1$ only, a current $I=(e^2/h)V_1$ flows from lead $1$ to lead $2$ 
but zero current flows into lead $3$.  We refer to this as a prefectly chiral 
conductance tensor.  Of course, if we reverse the sign of the flux then 
the chirality reverses with $V_1$ now inducing a current $(e^2/h)V_1$ from lead 
$1$ to lead $3$. 

Now consider the conductance in the interacting case, for the three fixed points 
that we have identified.  From the Kubo formula, we may write the DC linear conductance tensor as:
\be G_{jk}=\lim_{\omega \to 0}{-e^2\over h}{1\over \pi \omega L}\int_{-\infty}^\infty d\tau e^{i\omega \tau}
\int_0^LdxT<J_j(y,\tau )J_k(x,0)>,\label{GJJ}\ee
where $J_j=-i\partial_\tau \theta_j$. [At zero temperature, which I consider here, it is straightforward to take the 
zero frequency limit, in imaginary (Matsubara) formulation.  At finite $T$ it is neccessary to 
do an analytic continuation to real frequency first.]  We first transform from the $\phi_j$ 
basis to $\Phi_\mu$, inverting the transformation in Eq. (\ref{Phi}):
\be 
\phi_j=\sum_\mu v_{j\mu}\Phi_\mu .\ee
$\Phi_0$ makes no contribution to the conductance since $\Theta_0$ always 
obeys a D BC as discussed above. Therefore:
\be G_{jk}=\lim_{\omega \to 0}{-e^2\over h}{1\over \pi \omega L}
\sum_{\mu ,\nu = 1,2}v_{j\mu}v_{k\nu}\int_{-\infty}^\infty d\tau e^{i\omega \tau}
\int_0^LdxT<J_\mu (y,\tau )J_\nu (x,0)>,\ee
where $J_\mu = -i\partial_\tau \Theta_{\mu}$ and the result is independent of $y>0$. Here
\be v_{j\mu}=\sqrt{2/3}(\hat z \times \vec K_j)_\mu =-\sqrt{2/3}\sum_{\nu}\epsilon_{\mu \nu}K_{j}^{\nu}.\ee
To proceed we decompose:
\be \vec J = -i\partial_\tau \vec \Phi_{L}+i\partial_\tau \vec \Phi_{R}\equiv -\vec J_{L}+\vec J_R.\ee
The general type of BC of Eq. (\ref{bcRL}) allows us to regard the $\Phi_{R\mu} (x)$,s as the 
analytic continuation of the $\Phi_{L\nu}(x)$'s to the negative $x$ axis:
\be \vec \Phi_R(x)={\cal R}\vec \Phi_L(-x)+\vec C'\ee
and thus
\be \vec J_R(x)={\cal R}\vec J_L(-x).\ee
The Green's function for $\vec J_L$ is unaffected by the BC:
\be <J_{L\mu}(\tau+iy)J_{L\nu}(\tau '+ix)>={g\delta_{\mu \nu}\over 2[(\tau - \tau ')+i(y-x)]^2}.\ee
The $\tau$ integral in Eq. (\ref{GJJ}) gives:
\be \int_{-\infty}^\infty d\tau e^{i\omega \tau}
T<J_{L\mu} (y,\tau )J_{L\nu} (x,0)>=-2\pi \omega H(x-y)e^{\omega (y-x)}.\ee
Here $H(x)$ is the Heavyside step function, often written $\theta (x)$ but I avoid 
that notation here since $\theta (x)$ has another meaning. Thus we obtain:
\bea \int_{-\infty}^\infty d\tau e^{i\omega \tau}<J_\mu (\tau ,y)J_{\nu}(0,x)>
&=&-2\pi \omega [\delta_{\mu \nu}H(x-y)e^{\omega (y-x)}+\delta_{\mu \nu}H(y-x)e^{\omega (x-y)}
\nonumber \\
&&-{\cal R}_{\mu \nu}H(y+x)e^{-\omega (y+x)}-{\cal R}_{\nu \mu}H(-y-x)e^{\omega (y+x)}].\eea
Observing that $H(x-y)+H(y-x)=1$ and $H(x+y)=1$, $H(-x-y)=0$ since $x$ and $y$ are 
always positive in Eq. (\ref{GJJ}), we obtain:
\be G_{ij}=g{e^2\over h}\sum_{\mu ,\nu = 1,2}v_{j\mu}v_{k\nu}[\delta_{\mu \nu}-R_{\nu \mu}].
\label{GR}\ee

For the D BC on $\vec \Phi$, ${\cal R}=-I$ so:
\be G_{ij}=2g{e^2\over h}\vec v_j\cdot \vec v_k = 2g{e^2\over h}(\delta_{jk}-1/3),\ee
corresponding to $G_S=(4/3)g(e^2/h)$, $G_A=0$. We observed above that this is a stable fixed point for 
$g>3$, with $G_S>4e^2/h$. This exceeds the unitary bound on the conductance in the non-interacting case. 
An intuitive way of understanding why increasing $g$ leads to enhanced transmission is that 
attractive interactions can lead to pairing and either coherent pair tunnelling, or Andreev 
type processes (where an incident electron on one lead reflects as a hole while a pair 
is transmitted to a different lead) could lead to enhanced conductance. 

We can now also obtain the conductance tensor for the chiral fixed points, which are 
stable for $1<g<3$. Inserting Eqs. (\ref{Rxi} and (\ref{xi}) in (\ref{GR}) we obtain:
\bea G_S&=&{e^2\over h}{4g\over g^2+3}\nonumber \\
G_A&=& \pm {e^2\over h}{4g^2\over g^2+3}.\eea
For $g=1$ this reduces to the chiral conductance tensor discussed above in the 
non-interacting case with $G_S=\pm G_A=e^2/h$. However, for $g>1$ $G_A>G_S$ 
implying that a voltage on lead 1 not only leads to all current from 1 
flowing to 2 but some additional current also flows from 3 to 2. Intuitively, 
we might think that, as the electrons pass from lead 1 to 2 they attract 
some electrons from lead 3. If our hypothesis, discussed above is 
correct that the zero flux ``M'' fixed point is unstable, then 
presumably an infinitesimal flux could lead to an RG flow 
to these stable chiral fixed points. Such a device would 
have an interesting switching property.  Even a small magnetic  
field could switch the current completely from lead 2 to lead 3, 
at low enough temperatures and currents.

\section{Boundary Condition Changing Operators and the X-ray Edge Singularity}
There are some situations in condensed matter physics where we are interested 
in the response of a system to a sudden change in the Hamiltonian. A well-known 
example is the ``X-ray edge singularity'' in the adsorbtion intensity for 
X-rays in a metal, plotted versus X-ray energy. The X-ray dislodges an electron 
from a core level.  This is assumed to suddenly switch on a localized impurity potential 
which acts on the conduction electrons. Since I have argued that quite 
generally the low energy properties of quantum impurity problems 
are described by CIBC's we might expect that the low energy response 
to a sudden change in impurity interactions might be equivalent 
to the response to a sudden change in CIBC's.  Very fortunately, Cardy 
also developed a theory of BC changing operators which can be applied 
to this situation. In this lecture I will show how this theory 
 can be applied to the usual X-ray edge problem and to a multi-channel Kondo version.\cite{Affleck10}

\subsection{The X-Ray Edge Singularity}
When an X-ray is adsorbed by a metal it can raise an electron from a deep core level, 
several keV below the Fermi surface, up to the conduction band. Let $E_0$ be 
this large energy difference between the Fermi energy and the core level and 
let $\omega$ be the energy of the X-ray. At $T=0$, ignoring electron-electron 
interactions,  this transition is only possible for $\omega \geq \tilde E_0$. Here 
 $\tilde E_0$ is a ``renormalized'' value of $E_0$. 
I am assuming that the core level has a distinct energy, rather than itself 
being part of an energy band. This may be a reasonable approximation since 
core levels are assumed to be tightly bound to nuclei and to have very 
small tunnelling matrix elements to neighbouring nuclei.  Presumably 
the excited electron will eventually relax back to the core level, possibly 
emitting phonons or electron-hole pairs.  This is ignored in the usual treatment 
of X-ray edge singularities. Thus we are effectively ignoring the 
finite width of the excited electron states. Thus the X-ray adsorption 
intensity, $I(\omega )$ will be strictly zero for $\omega \leq \tilde E_0$, in this approximation. 
When the core electron is excited into the conduction band, it leaves behind a core 
hole, which interacts with all the electrons in the conduction band. Note 
that the only interaction being considered here is the one between the 
core hole and the conduction electrons. 
The X-ray edge singularity, at $\omega =\tilde E_0$, in this approximation, is 
determined by the response of the conduction electrons to the sudden 
appearance of the core hole potential, at the instant that the 
X-ray is adsorbed. It turns out that, for $\omega$ only 
slightly larger then $\tilde E_0$, very close to the threshold, 
$I(\omega )$ is determined only by the conduction electron states very 
close to $\epsilon_F$; this fact allows us to apply 
low energy effective Hamiltonian methods. The difference between $\tilde E_0$ and $E_0$ 
arises from the energy shift of the filled Fermi sea due to the core hole potential. 
Not including the interaction with the external 
electromagnetic field, which I turn to momentarily, the Hamiltonian is simply:
\be H=\int d^3r\left[\psi^{\dagger}(\vec r)\left(-{\nabla^2\over 2m}-\epsilon_F \right) \psi (\vec r)
+\tilde Vbb^\dagger \delta^3(r)\psi^\dagger \psi \right]+E_0b^\dagger b ,\label{HXR}\ee
where $b$ annihilates an electron in the core level at $\vec r =0$. I have assumed, for simplicity, 
that the core hole potential, $\tilde V\delta^3(r)$,  is a spherically symmetric $\delta$-function. These 
assumptions can be easily relaxed. 
Following the same steps as in Sec. I, 
we can reduce the problem to a one-dimensional one, with left-movers only:
\be H= {1\over 2\pi }i\int_{-\infty}^\infty dr \psi_L^\dagger {d\over dr}\psi_L
+{V\over 2\pi}bb^\dagger \psi_L^\dagger (0)\psi_L(0) +E_0b^\dagger b.
\label{HXL}\ee
(I have set $v_F=1$ and $V\propto \tilde V$.) It is convenient to bosonize.  We may introduce a left-moving boson 
only:
\be \psi_L\propto e^{i\sqrt{4\pi}\phi_L},\ee
\be H=\int_{-\infty}^\infty (\partial_x\phi_L)^2-{V\over \sqrt{\pi}}bb^\dagger \partial_x\phi_L+E_0bb^\dagger.\ee
The solubility of this model hinges on the fact that $b^\dagger b$ commutes with $H$. Thus the Hilbert 
Space breaks up into two parts, in which the core level is either empty or occupied. The spectrum of 
the Hamiltonian, in each sector of the Hilbert Space, is basically trivial.  In the sector where the 
core level is occupied, $b^\dagger b=1$, we get the spectrum of free electrons with no impurity:
\be H_0=\int_{-\infty}^\infty (\partial_x\phi_L)^2.\ee
  In the 
sector with $b^\dagger b=0$ we get the spectrum with a potential scatterer present:
\be H_1=\int_{-\infty}^\infty (\partial_x\phi_L)^2-{V\over \sqrt{\pi}} \partial_x\phi_L+E_0.\ee
What makes this problem somewhat non-trivial is that, to obtain the edge singularity, 
we must calculate the Green's function of the operator which couples to the electromagnetic 
field associated with the X-rays: $\psi^\dagger (t,r=0)b(t)$. This operator, which excites an electron 
from the core level into the conduction band, mixes the 2 sectors 
of the Hilbert Space. There is also some interest in calculating the Green's function of 
the operator $b(t)$ itself; this is associated with photo-emission processes in which the 
core electron is ejected from the metal by the X-ray. Again this operator mixes the 
two sectors of the Hilbert space. 

In what may have been the first paper on bosonization, in 1969, Schotte and Schotte \cite{Schotte} observed 
that these Green's functions can be calculated by taking advantage of the fact that the two 
Hamiltonians, $H_0$ and $H_1$ are related by a canoical transformation (and a shift 
of the ground state energy). To see this note that we may write $H_1$ in the form:
\be H_1=\int_{-\infty}^\infty (\partial_x\tilde \phi_L)^2+\hbox{constant},\ee
where:
\be \tilde \phi_L(x)\equiv \phi_L(x) -{V\over 4\sqrt{\pi}}\hbox{sgn}(x).\label{phit}\ee
Using the commutator:
\be [{\partial_y\phi_L(y)},\phi_L(y)]={-i\over 2}\delta (x-y)],\ee
we see that:
\be H_1=U^\dagger H_0U+\hbox{constant},\label{U}\ee
with the canonical transformation:
\be U=\exp [-iV\phi_L(0)/\sqrt{\pi}].\label{Uphi}\ee
($U$ can only be considered a unitary operator if we work in the extended Hilbert space 
which includes states with {\it all} possible BCs. $U$ maps whole sectors on the 
Hilbert space, with particular BCs, into each other.)
Consider the core electron Green's function, $<b(t)^\dagger b(0)>$. We may 
write:
\be b^\dagger (t)=e^{iHt}b^\dagger e^{-iHt}=e^{iH_0t}b^\dagger e^{-iH_1t}.\ee
This is valid because, due to Fermi statistics,  the core level must be vacant before $b^\dagger$ acts, 
and occupied after it acts. i.e. we can replace the $bb^\dagger$ factor in $H$ by $1$ on 
the right hand side and by $0$ on the left. But $H_0$ and $H_1$ commute with $b$ so we have:
\be <0|b^\dagger (t)b(0)|0>=<1|b^\dagger b|1><\tilde 0|e^{iH_0t}e^{-iH_1t}|\tilde 0>.\ee
Here $|0>$ is the ground state of the system, including the core level and 
the conduction electrons.  This state can be written: $|0>=|1>|\tilde 0>$ where 
$|1>$ is the state with the core level occupied and $|\tilde 0>$ is the filled Fermi sea ground state 
of the conduction electrons, with no impurity potential. Using Eq. (\ref{U}) we see that:
\be <\tilde 0||e^{iH_0t}e^{-iH_1t}|\tilde 0>=e^{-i\tilde E_0t}<\tilde 0|e^{iH_0t}U^\dagger e^{-iH_0t}U|\tilde 0>
=e^{-i\tilde E_0t}<\tilde 0|U(t)^\dagger U(0)|\tilde 0>
.\ee
Using our explicit expression, Eq. (\ref{Uphi}) for $U$, we have reduced the calculation 
to one involving only a free boson Green's function:
\be <0|b^\dagger (t)b(0)|0>= e^{-i\tilde E_0t}<e^{iV\phi_L(t,0)/\sqrt{\pi}}e^{-iV\phi_L(0,0)/\sqrt{\pi}}>
\propto { e^{-i\tilde E_0t}\over t^{V^2/4\pi^2}}.\ee
Similarly, to get the Green's function of $b^\dagger (t)\psi_L (t,0)$ I use:
\be b^\dagger (t)\psi_L(t,0)=e^{iH_0t}b^\dagger e^{-iH_1t}\psi_L(t,0)\ee
and thus
\bea <0|b^\dagger (t)\psi_L(t,0)b(0)\psi_L(0,0)|0>&=&e^{-i\tilde E_0t}<\tilde 0|U^\dagger (t)U(0)\psi_L(t,0)\psi_L^\dagger(0,0)|\tilde 0>\nonumber \\
&\propto& e^{-i\tilde E_0t}<e^{i\sqrt{4\pi}(1+V/2\pi )\phi_L(t,0)}e^{-i\sqrt{4\pi}(1+V/2\pi )\phi_L(0,0)}>
\propto {e^{-i\tilde E_0t}\over t^{(1+V/2\pi )^2}}.\eea
Finally, we Fourier transform to get the X-ray edge singularity:
\be \int_{-\infty}^\infty dt e^{i\omega t} <0|b^\dagger (t)\psi_L(t,0)b(0)\psi_L^\dagger (0,0)|0>
\propto \int_{-\infty}^\infty dt {e^{i(\omega -\tilde E_0)t}\over (t-i\delta )^{(1+V/2\pi )^2}}
\propto {\theta (\omega -\tilde E_0)\over (\omega -\tilde E_0)^{1-(1+V/2\pi )^2}}.\ee

This result is conventionally written in terms of a phase shift at the Fermi surface, rather than 
the potential strength, $V$. The connection can be readily seen from Eq. (\ref{phit}) and the 
bosonization formula:
\be \psi_L(x)\propto e^{i\sqrt{4\pi}\phi_L(x)}e^{iV\cdot \hbox{sgn}(x)/2}.\ee
Since $\psi_L(x)$ for $x<0$ represents the outgoing field, we see that:
\be \psi_{\hbox{out}}=e^{2i\delta}\psi_{\hbox{in}},\ee
where the phase shift is:
\be \delta = -V/2.\ee
In fact, the parameter $V$ appearing in the bosonized Hamiltonian should be regarded as a renormalized one. 
Its physical meaning is the phase shift at the Fermi surface induced by the core hole. Only for 
small $\tilde V$ is it linearly related to the bare potential. Even if the core hole potential 
has a finite range, we expect the formulas for the X-ray edge singularity to still be correct, 
when expressed in terms of the phase shift at the Fermi surface, $\delta$. More generally, 
if the core hole potential is not a $\delta$-function, but is still spherically symmetric, 
a similar expression arises for the X-ray edge singularity with the exponent involving a 
sum over phase shifts at the Fermi surface, $\delta_{l}$, in all angular momentum channels, $l$. 

The connection with a boundary condition changing operator (BCCO)\cite{CardyH,Cardy1}
 is now fairly evident.
If we revert to the formulation of the model on the semi-infinite line, $r>0$, then 
the boundary condition is:
\be \psi_R(0)=e^{2i\delta}\psi_L(0).\ee
The operator, $U$ or $b$, which creates the core hole potential in the Hamiltonian can be 
viewed as changing the boundary condition, by changing the phase shift $\delta$. 
It is interesting to consider the relationship between the finite size spectrum 
with various BC's and the scaling dimensions of $b$ and $\psi_L(0)^\dagger b$. 
We consider the system on a line of length $l$ with a fixed BC $\psi_R(l)=-\psi_L(l)$ 
at the far end. Equivalently, in the purely left-moving formulation, for $\delta =0$ 
we have anti-periodic BC's on a circle of circumference $2l$: $\psi_L(x+2l)=-\psi_L(x)$. 
This corresponds to periodic BC's on the left-moving boson field, 
\be \phi_L(x+2l)=\phi_L(x)+\sqrt{\pi Q},\ \  (Q=0,\pm 1,\pm 2, \ldots ).\ee
The mode expansion is:
\be \phi_L(t,x)=\sqrt{\pi}{(t+x)\over 2l}Q+\sum_{m=1}^\infty {1\over \sqrt{2\pi m}}
\left[ \exp (-i\pi m(t+x)/l)a_m+h.c.\right].\label{mode}\ee
The finite size spectrum is:
\be E=\int_{-l}^l(\partial_x\phi_L)^2={\pi \over l}\left[-{1\over 24}+{1\over 2}Q^2+\sum_{m=1}^\infty mn_m\right].\ee
The universal ground state energy term, $-\pi /(24 l)$ has been included. We see that $Q$ can be 
identified with the charge of the state (measured relative to the filled Fermi sea). There 
is a one to one correspondance between the states in the FSS with excitation energy $\pi x/l$ and 
operators with dimension $x$ in the free fermion theory. For example, $Q=\pm 1$ corresponds 
to $\psi_L^\dagger$ and $\psi_L$ respectively, of dimension $x=1/2$. If we impose the 
``same'' BC at both ends:
\bea \psi_R(0)&=&e^{2i\delta}\psi_L(0)\nonumber \\
\psi_R(l)&=&-e^{2i\delta}\psi_L(l)\eea 
then this corresponds to the same anti-periodic BC on $\psi_L(x)$ in the purely left-moving 
formulation, so the spectrum is unchanged.  On the other hand if the phase shift is 
inserted at $x=0$ only, then $\psi_L(x+2l)=-e^{-2i\delta}\psi_L(x)$, corresponding to:
\be \phi_L(x+2l)-\phi_L(0)=\sqrt{\pi}(n-\delta /\pi ),\ee
corresponding to the replacement $Q\to Q-\delta /\pi$ in Eq. (\ref{mode}). 
The FSS is now modified to:
\be E=\int_{-l}^l(\partial_x\phi_L)^2={\pi \over l}\left[-{1\over 24}+{1\over 2}(Q-\delta /\pi )^2+\sum_{m=1}^\infty mn_m\right].\ee
In particular, the change in ground state energy due to the phase shift is:
\be E_0(\delta )-E_0(0)={\pi \over l}{1\over 2}\left({\delta\over \pi}\right)^2.\label{Edel}\ee
Actually, adding the potential scattering also changes the ground state energy 
by a non-universal term, of O(1) which was adsorbed into $\tilde E_0$ in the 
above discussion. It is the term of O($1/l)$ which is universal and determines 
scaling exponents. 
The corresponding scaling dimension, $x=(\delta /\pi )^2/2$ is precisely the scaling 
dimension of the BCCO $b$ (or $U$). The energy of the excited state with $Q=1$ obeys:
\be E (Q,\delta )-E_0(0)={\pi \over l}{1\over 2}\left(1-{\delta\over \pi}\right)^2.\ee
The corresponding scaling dimension, $x=(1-\delta /\pi )^2/2$, is the scaling 
dimension of $\psi_L(0)^\dagger b$.

\begin{figure}[th]
\begin{center}
\includegraphics[clip,width=8cm]{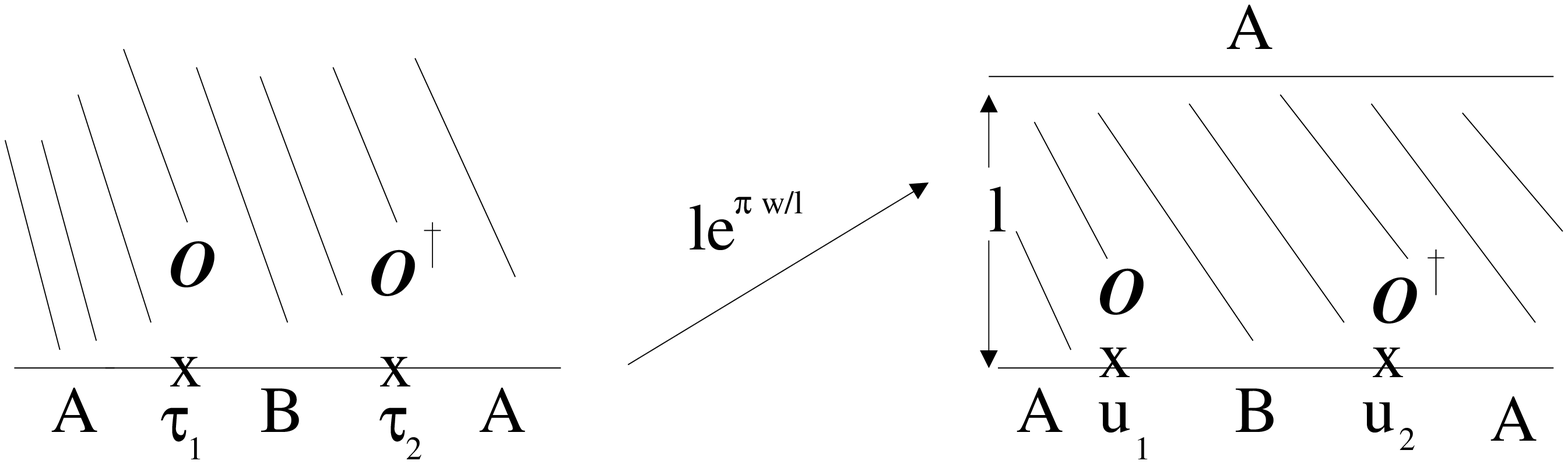}
\caption{BCCO's act at times $\tau_1$ and $\tau_2$ on the semi-infinite plane. This 
is conformally mapped to the infinite strip with the BCCO's acting on the lower boundary.}
\label{fig:BCCO}
\end{center}
\end{figure}

This is all in accord with Cardy's general theory of BCCO's.  An operator which 
changes the BC's from $A$ to $B$ generally has a scaling dimension, $x$, which gives 
the ground state energy on a finite strip of length $l$ with BC's $A$ at 
one end of the strip and $B$ at the other end, measured relative to 
the absolute ground state energy with the same BC's $A$, at both ends. This follows 
by making a conformal transformation from the semi-infinite plane to the finite strip. 
Explicity, consider acting with a primary BCCO ${\cal O}$ at time $\tau_1$ at the edge, $x=0$ 
of a semi-infinite plane.  Assume ${\cal O}$ changes the BC's from $A$ to $B$.  Then 
at time $\tau_2$ change the BC's back to $A$ with the hermitean conjugate operator 
${\cal O}^\dagger$, also acting at $x=0$, as shown in Fig. (\ref{fig:BCCO}). Then 
the Green's function on the semi-infinite plane is:
\be <A|{\cal O}(\tau_1){\cal O}^\dagger (\tau_2)|A>={1\over (\tau_2-\tau_1)^{2x}},\ee
where $x$ is the scaling dimension of ${\cal O}$. Now we make a conformal transformation 
from the semi-infinite plane, $z=\tau +ix$, ($x\geq 0$) to the finite width strip:
$w=u+iv$, $0<v<l$:
\be z=le^{\pi w/l}.\ee
Note that the positive real axis, $x=0$, $\tau >0$, maps onto the bottom of the strip, $v=0$. 
Choosing $\tau_1$, $\tau_2>0$, both points map onto the bottom of the strip. Note that, 
on the strip, the BC's are $A$ at all times at the upper boundary, $v=l$ but change 
from $A$ to $B$ and then back to $A$ on the lower strip, at times $u_1$, $u_2$. 
The Green's function on the strip is given by:
\be <AA|{\cal O}(u_1){\cal O}^\dagger (u_2)|AA>=\left({\pi \over 2l\sinh [(\pi (u_1-u_2)/(2l)]}\right)^{2x}.
\label{OOl}\ee
Here $|AA>$ denotes the ground state of the system on the strip of length $l$ with the same BC, $A$, 
at both ends of the strip. We may insert a complete set of states:
\be  <AA|{\cal O}(u_1){\cal O}^\dagger (u_2)|AA>=\sum_n|<AA|{\cal O}|n>|^2e^{-E_n(u_2-u_1)}.\label{OOs}\ee
However the states $|n>$ must all be states in the Hilbert Space with {\it different} BC's $A$ at $v=0$ 
and $B$ at $v=l$. The corresponding energies, $E_n$ are the energies of the states with 
these different BC's measured relative to the absolute ground state energy with the same 
BC's $A$ at both ends. Taking the limit of large $u_2-u_1$ in Eqs. (\ref{OOl}) and (\ref{OOs}) 
we see that 
\be {\pi x\over l}=(E^{{\cal O}}_{AB}-E_0),\ee
where $E^{{\cal O}}_{AB}$ is the lowest energy state with BC's $A$ and $B$ produced by 
the primary operator ${\cal O}$. The lowest dimension BCCO will simply produce 
the ground state with BC's $A$ and $B$. This corresponds to the operator $b$ in 
the X-ray edge model. On the other hand, the lowest energy state produced could 
be an excited state with BC's $A$ and $B$ as in the example of $b\psi_L^\dagger (0)$, 
which produces the primary excited state with $Q=1$ and BC's twisted by the phase $\delta$.
 
\subsection{X-ray Edge Singularites and the Kondo Model}
This BCCO approach has various other applications \cite{Affleck10} that go beyond the Schotte and Schotte results. 
One of them is to the Kondo model. So far we have been ignoring electron spin. If the 
core level is doubly occupied in the ground state, then it would have spin-1/2 
after one electron is ejected from it by the X-ray. An initially spin-0 ion 
would then acquire a net spin-1/2.  In addition to the potential-scattering 
interaction with the localized core hole, there would also be a  Kondo interaction. 
Thus, it is interesting to consider the effect of suddenly turning on a Kondo 
interaction. We might expect that the Kondo effect could dominate the X-ray 
edge exponent, at least at low enough temperatures $T\ll T_K$ and frequencies:
$\omega-\tilde E_0\ll T_K$. Thus, we consider the Hamiltonian:
\be H= {i\over 2\pi }\int_{-\infty}^\infty dr \psi_L^{i\alpha \dagger} {d\over dr}\psi_{Li\alpha}
+\lambda \psi_L^{i\gamma\dagger} (0){\vec \sigma_\gamma^\delta \over 2}\psi_{Li\delta}(0)\cdot 
b^{\alpha \dagger}{\vec \sigma_\alpha^\beta \over 2}b_\beta +E_0b^{\alpha \dagger}b_\alpha  .
\label{KX}\ee
Now $b^{\alpha \dagger}$ creates a core electron with spin $\alpha$ and we have considered 
the general case of $k$ channels of conduction electrons, $i=i$, $2$, $3$, $\ldots$  
Again we are interested in Green's functions for the core electron operator, $b_\alpha (t)$ 
and also the operators $b^{\alpha \dagger}(t)\psi_{Li\beta}(t,0)$.  These Green's functions 
should exhibit a non-trivial cross-over with frequency, or time, but at long times 
(frequencies very close to the threshold) we expect to be able to calculate 
them using properties of the Kondo fixed point. Since the operator $b_{\alpha}$ 
creates the impurity spin, thus turning on the Kondo effect, it is again a BCCO. 
In this case, it should switch the BC from free to Kondo. Thus we expect 
its infrared scaling dimension to be given by the energy of the ground 
state with a free BC at one end of the finite system and a Kondo BC at the other. 
This spectrum is given by fusion with the $j=1/2 $ primary in the spin sector. 
The ground state with these BC's is always the $j=1/2$ primary itself, 
of dimension 
\be x=(3/4)/(2+k).\label{Xxk}\ee

  This follows because the free spectrum 
includes the charge zero, spin $j=0$ flavour singlet, $(0,0,I)$. Fusion with the 
spin $j$ primary always gives $0,j,I)$ among other operators. This appears 
to have the lowest dimension of all fusion products. 
Note that this operator {\it does not} occur 
in the operator spectrum considered earlier at the Kondo fixed point.  There 
we only considered operators produced by double fusion, corresponding to the FSS 
with Kondo BC's at both ends of the finite system. This gives the operator 
spectrum with a fixed, Kondo BC.  But for the Kondo X-ray problem, we must 
consider the corresponding BCCO. In general, for a CIBC obtained by 
fusion with some operator ${\cal O}$ from a non-interacting BC, we 
may expect that the BCCO will be ${\cal O}$ itself. We may check this result 
by a more elementary method in the single channel case. There the Kondo 
fixed point is equivalent to a phase shift of magnitude $\pi /2$ for 
both spin up and spin down, $\delta_{\uparrow ,\downarrow}$. The energy of this state, 
from Eq. (\ref{Edel})  
is simply $(1/2\pi l)[(\delta_{\uparrow}/\pi )^2+(\delta_{\downarrow}/\pi )^2]$ implying a dimension
$x=1/4$. This agrees with Eq. (\ref{Xxk}) in the special case $k=1$. 
We may also consider the dimensions of the operators $\psi^{i\alpha \dagger}_L(0)b_\beta $. 
This operator has $Q=1$ (one extra electron added to the conduction band), transforms 
under the fundamental representation of flavour, of dimension $k$ and has spin either $j=0$ or $j=1$ 
depending on how we sum over the spin indices $\alpha$ and $\beta$. The free 
spectrum always contains the operator corresponding to the fermion field itself, 
$(Q=1,j=1/2,k)$ (where $k$ now denotes the $k$-dimensional fundamental representation 
of $SU(k)$) and fusion with $j=1/2$ gives $(Q=1,j=0,k)$ for all $k$ and $(Q=1,j=1,k)$ 
for $k\geq 2$. These operators have dimension:
\be x_j={1\over 4k}+{k^2-1\over 2k(2+k)}+{j(j+1)\over 2+k}.\ \  (j=0,1)
\label{Xxj}\ee
(It can be seen than $x_{1/2}=1/2$ corresponding to the free fermion operator.) 
Again we may check the case $k=1$ by more elementary arguments. We may 
find the unitary operators corresponding to $b_\alpha$ as:
\be b_\alpha \propto \exp [2i(\delta_\uparrow \phi_{\uparrow L}+\delta_\downarrow \phi_{\downarrow L}],\ee
where we have introduced separate bosons for spin up and spin down electrons. $\delta_\uparrow$, 
$\delta_{\downarrow}$ can depend on $\alpha$.  It is convenient 
to switch to charge and spin bosons,
\be \phi_{c/s}\equiv {\phi_{\uparrow}\pm \phi_{\downarrow}\over \sqrt{2}}.\ee 
By choosing $(\delta_\uparrow ,\delta_\downarrow)=(\pi /2,-\pi /2)$ for $b_\uparrow$ 
and $(\delta_\uparrow ,\delta_\downarrow )=(-\pi /2,\pi /2)$ for $b_{\downarrow}$ we obtain:
\be b_{\uparrow /\downarrow}\propto \exp (\pm i\sqrt{2\pi}\phi_{Ls}).\ee
These have the correct $S^z$ quantum numbers as can be seen by comparing with the 
standard bosonization formula for $\psi_{L\alpha}$:
\be \psi_{\uparrow /\downarrow L}\propto \exp (i\sqrt{2\pi} \phi_{Lc})\exp (\pm i\sqrt{2\pi}\phi_{Ls}).\ee
$\exp (\pm i\sqrt{2\pi}\phi_{Ls})$ can be identified with $g_{L\uparrow/\downarrow}$ 
the chiral component of the WZW model fundamental field. 
We then see that the spin singlet operator, $\exp (i\sqrt{2\pi} \phi_{Lc})$, 
has $x=1/4$ in agreement with Eq. (\ref{Xxj}).  On the other hand, 
the triplet operators have dimension $x=5/4$.  Since there is 
no $j=1$ primary for $k=1$ they contain Kac-Moody descendents, i.e. 
the spin current operator.

\section{Conclusions}
Apart from the examples discussed in these lectures, BCFT techniques have 
been applied to a number of other quantum impurity problems, including 
the following.  We can consider a local cluster of impurities.  At 
distances large compared to the separation between the impurities the 
same methods can be applied.  The 2-impurity Kondo model exhibits a NFL
fixed point which can be obtained\cite{Affleck8} by a conformal embedding which includes 
an Ising sector in which the fusion is performed. The 3-impurity 
Kondo model also exhibits a novel NFL fixed point.  It was obtained\cite{Ingersent} 
by a different conformal embedding with fusion in a $Z_8$ parafermion 
CFT sector. Impurities in $SU(3)$ spin chains\cite{Affleck12} and quantum Brownian 
motion\cite{Affleck11} were also solved by these techniques.  They were even 
applied\cite{Affleck9} to a high energy physics model associated with Callan and Rubukaov. 
This describes a super-heavy magnetic monopole interacting with $k$-flavours of
 effectively massless fermions (quarks and leptons).  The monopole is 
actually a dyon having a set of electric charge states as well as a 
magnetic charge. When the fermions scatter off the dyon they can 
exchange electric charge. In this case fusion takes place in the charge sector. 

The assumption that essentially arbitrary impurity interactions, possibly involving
localized impurity degrees of freedom, interacting with a gapless continuum, 
renormalizes at low energies to a CIBC 
has worked in numerous examples. It appears to be generally valid and will 
likely find many other applications in the future. 
\section{Ackowledgements}
I would like to thank my collaborators in the work discussed here, including:
Andreas Ludwig, Sebastian Eggert, Masaki Oshikawa, Claudio Chamon, Ming-Shyang Chang, Erik Sorensen and 
Nicolas Laflorencie.

\end{document}